# HenTwin: A Multimodal Digital Twin Framework for Longitudinal Biological State Monitoring in Laying Hens

Yashan Dhaliwal, *Undergraduate Student Member, IEEE*, Shreya Rao, *Graduate Student Member, IEEE* and Suresh Neethirajan, *Member, IEEE*

***Abstract*— Early-life monitoring in laying hens remains constrained by fragmented single-modality sensing and the absence of formal system-level state representations. HenTwin, a multimodal digital twin framework implemented as a five-layer IoT architecture, formalizes flock-level multimodal biological state dynamics from hatch through 25 weeks of age. A four-dimensional biological state vector integrating body surface temperature, acoustic energy entropy, band energy ratio, and optical-flow-based motion is defined, with the temperature–humidity index treated as an exogenous environmental input to preserve intervention capability. A discrete-time state transition model is estimated from 25 weeks of longitudinal multimodal data collected from 150 Lohmann LSL-Lite hens across five controlled rooms at the Atlantic Poultry Research Centre, Dalhousie University. The estimated transition matrix exhibits modality-specific persistence while remaining asymptotically stable. Perturbation analysis demonstrates that a sustained +2.0 THI increase produces a stable long-run acoustic entropy elevation of 0.54 nats, approximately one-quarter of the entire 1.87-nat developmental decline observed across the study period. Pettitt change-point detection identifies coordinated multimodal developmental state transitions at Weeks 12–14. Cross-room validation suggests that structural transition parameters are partially transferable across rooms, whereas environmental input sensitivity requires room-specific calibration, supporting a two-tier IoT deployment architecture. Leave-one-out cross-validation demonstrates consistent out-of-sample model performance. HenTwin takes a first step toward formal, state-aware digital twin inference in precision livestock farming.**

## I. Introduction

Early-life development in laying hens, spanning hatch through approximately 25 weeks, constitutes a critical biological window that shapes long-term health, productivity, and behavioural stability [1]. During this period, physiological systems governing thermoregulation, vocal communication, and locomotor coordination undergo rapid maturation, and disruptions to welfare-relevant conditions, whether arising from thermal stress, social perturbation, or management events, may have lasting consequences for flock performance [1,2]. Despite this recognized importance, monitoring practices in both research and commercial settings continue to rely predominantly on periodic manual inspection, subjective scoring protocols, and single-modality sensors that provide fragmented, low-resolution snapshots of animal condition [3]. The result is a mismatch between the complexity of the biological system and the coarseness of the infrastructure used to monitor it.

Precision Livestock Farming (PLF) has expanded the use of continuous sensing technologies in poultry production, including thermal imaging, acoustics, machine vision, inertial measurement, and environmental instrumentation [4,5]. However, these signals are typically analyzed independently rather than integrated into a coherent system-level representation [4,6]. Individual modality outputs, such as temperature measurements, acoustic descriptors, or activity indices, represent different projections of the same underlying biological process, yet no formal framework currently exists to fuse these observations into a unified, temporally evolving state estimate. Without such formalization, PLF systems remain largely descriptive rather than inferential; they characterize what has been observed but provide limited capability to determine what biological state the flock occupies, how that state evolves over time, or how it may respond to environmental perturbations [6,7].

Digital twin (DT) technology offers a principled framework for addressing this limitation. A digital twin is a synchronized virtual representation of a physical entity that encapsulates both current state and dynamic evolution laws, enabling monitoring, simulation, forecasting, and intervention-aware decision support [8]. Digital twin principles have gained traction across

Y. Dhaliwal is with the Faculty of Agriculture, Dalhousie University, Truro, NS B2N 5E3, Canada (e-mail: yashan.dhaliwal@dal.ca).

S. Rao is with the Faculty of Computer Science, Dalhousie University, Halifax, NS B3H 4R2, Canada (e-mail: sh316686@dal.ca).

S. Neethirajan is with the Faculty of Agriculture, Dalhousie University, Truro, NS B2N 5E3, Canada, and with the Faculty of Computer Science, Dalhousie University, Halifax, NS B3H 4R2, Canada (Corresponding author: S. Neethirajan; e-mail: sureshraja@dal.ca).

manufacturing, infrastructure health monitoring, and smart agriculture, yet their application to biological systems remains comparatively immature [6,7]. In PLF specifically, the gap between DT potential and formal implementation is well documented, with major barriers including undefined state spaces, absent transition dynamics, and limited predictive validation [6,7]. Many existing livestock DT implementations function primarily as sensor dashboards with aggregated outputs, lacking the state-transition models and predictive capabilities required for true digital twin functionality [6,7].

HenTwin, a multimodal digital twin for early-life laying hens, is implemented as a five-layer IoT architecture spanning sensing, edge processing, state-space modelling, twin abstraction, and decision support. It is grounded in a 25-week longitudinal dataset from 150 Lohmann LSL-Lite hens across five controlled environmental rooms at the Atlantic Poultry Research Centre, Dalhousie University, with Rooms 1 and 5 as the primary analytical units. HenTwin is formulated as a multimodal biological state framework in which thermal, acoustic, behavioural, and environmental signals are integrated into a unified state representation that supports monitoring, simulation, and digital twin inference in early-life laying hens. The core contributions of this work are:

1. A four-dimensional biological state vector S(t) integrating thermal, acoustic, and behavioural modalities, with THI incorporated as an exogenous forcing input to preserve intervention capability.
2. A discrete-time state transition framework comprising a diagonal persistence matrix A and environmental input coupling vector B, estimated from 25 weeks of multimodal longitudinal observations.
3. Validation of model behaviour through leave-one-out cross-validation, stability analysis, and perturbation simulation, demonstrating that sustained THI increases produce measurable shifts in biological state trajectories.
4. A two-tier generalisability architecture showing that structural transition dynamics are partially transferable across rooms while environmental sensitivity requires room-specific calibration, providing practical guidance for multi-barn IoT deployment.

The remainder of this paper is organized as follows: Section II reviews related work, Section III describes the experimental setup and sensor suite, Section IV details data processing and feature extraction, Section V formalizes the HenTwin state-space framework, Section VI presents statistical and validation analyses; Section VII reports empirical results; Section VIII evaluates state transition analysis and cross-room validation; Section IX discusses implications and limitations, Section X concludes.

## II. Related work

### *A. Precision Livestock Farming and Multimodal Welfare Sensing*

Precision livestock farming has progressively expanded the role of automated sensing in poultry production, transitioning from periodic human inspection toward continuous, data-driven monitoring. Early systems were single modality, capturing one dimension of flock state in isolation. Herborn et al. showed that spectral entropy of early-life distress calls correlates strongly with manual welfare assessments, establishing it as a non-invasive welfare indicator [9]. Soster et al. further demonstrated that vocalization patterns in broilers shift measurably with age, time of day, and environmental conditions including heat stress, confirming the sensitivity of acoustic signals to early-life welfare-relevant stimuli [2]. Building on this, Manikandan and Neethirajan extended this by showing that energy entropy and band energy ratio provide interpretable acoustic signals of flock stress state [5]. Parallel work in thermal modalities established that infrared thermography reliably tracks body surface temperature responses to environmental thermal load, with systematic reviews confirming the sensitivity of specific anatomical regions to stress-induced vasomotor changes [10].

Behavioural monitoring through computer vision has also advanced. Dawkins et al. established that optical flow patterns from overhead cameras correlate with key welfare outcomes including mortality, hockburn, and gait scores in commercial broiler flocks. The research demonstrated that group-level motion statistics carry enough welfare information without requiring individual bird identification [11]. More recently, a systematic review of AI-powered vocalization analysis in poultry synthesised evidence across acoustic feature extraction methods from MFCCs and spectral entropy to deep learning architectures. This further confirms the expansion of non-invasive welfare indicators available from audio streams [12].

A pilot study directly preceding the present work demonstrated the technical feasibility of parallel multimodal data acquisition across thermal, acoustic, optical-flow, and environmental modalities in early-life laying hens. The study characterised age-related developmental trajectories and cross-modal associations across a 20-week rearing period [13]. The study explicitly scoped itself as descriptive and did not state welfare state classification, causal inference, or predictive modelling. However, it established the multimodal sensing infrastructure and confirmed cross-modal coupling as the empirical foundation on which the present HenTwin framework builds.

Despite this accumulation of unimodal and multimodal evidence, signals from different modalities are almost universally analysed in isolation or fused post-hoc at the feature level, without integration into a formal state representation. Li et al. observed in their review of automated welfare monitoring in broilers and laying hens that multimodal integration is a necessary future direction but remains unrealised in practice [4]. A recent multimodal AI review covering 130 studies on laying hen welfare confirmed this pattern, noting that unimodal models dominate the literature and that multimodal fusion is recognised as crucial yet remains underspecified [14]. The result is a landscape of aggregation pipelines rather than state models: systems that report what individual sensors measure but cannot represent what state the flock is in, how that state evolves, or how it would respond to intervention.

### B. Digital Twin Frameworks in Agriculture and Livestock

Digital twin technology has attracted substantial attention in agricultural and livestock contexts as a framework for moving beyond descriptive sensing toward dynamic, model-driven management. Garro and Sorrenti formalised a hybrid digital twin architecture aligned with the ISO 23247 standard. They defined a reference architecture that integrates physics-based and data-driven models across a layered architecture comprising physical, communication, digital twin, and user entities [8]. Drawing on this framework [8], a digital twin is distinguished from a monitoring dashboard by three properties: a synchronized state representation, an explicit dynamic evolution model, and bidirectional predictive or intervention-aware capacity.

Reviews of digital twin adoption in agriculture document a growing range of applications spanning crop management, environmental control, and livestock health monitoring [15]. A multi-layer digital twin framework for smart agriculture and livestock, based on systematic review of publications from 2021 to 2025. It identified scalability and interoperability as high-impact implementation barriers while demonstrating that microservice-based architectures can substantially improve latency and transparency metrics [16]. In livestock supply chain management, recent synthesis work situates farm-level cyber-physical capabilities within chain-level traceability and policy compliance requirements, confirming the broader operational relevance of formal digital twin architectures beyond single-facility deployment [17].

Within precision livestock farming specifically, the gap between digital twin rhetoric and formal implementation has been documented explicitly. Neethirajan identified that existing work labelled as livestock digital twins largely consists of sensor dashboards and aggregated KPI outputs, lacking the state-transition machinery and predictive capacity that the ISO 23247 definition requires [6,7]. An earlier perspectives paper by Neethirajan and Kemp established the conceptual case for digital twins in livestock farming but did not define a formal state vector, estimate transition dynamics, or validate predictive capacity [18]. To the best of our knowledge, no published study has simultaneously defined an explicit multimodal biological state vector for laying hens, estimated a discrete-time transition model from longitudinal multimodal observations, and demonstrated forward simulation of state response to environmental perturbation within a digital twin framework. HenTwin addresses this gap by implementing all three definitional criteria within a formally estimated and empirically validated framework.

### C. State-Space Modeling in Biological Systems

State-space models (SSMs) provide the mathematical infrastructure for the HenTwin transition formulation, representing a system through a latent state that evolves according to a transition equation and is observed through noisy measurements. SSMs have a well-established history in ecological time-series analysis, where the Kalman filter enables efficient inference in linear Gaussian formulations and extensions including the extended Kalman filter handle nonlinear cases [19]. Patterson et al. applied hidden Markov models to infer discrete behavioural states from continuous animal tracking data, demonstrating SSM applicability in biological systems [20]. Kalman filter approaches have been applied to dairy milk yield and oestrus detection, confirming recursive state estimation feasibility from sparse biological time series. [21].

The linear discrete-time formulation adopted in HenTwin belongs to the class of linear Gaussian SSMs tractable with ordinary least squares estimation with the environmental input treated as exogenous to enable counterfactual simulation [19].

TABLE I
COMPARISON OF HENTWIN WITH CLOSEST PRIOR WORK ACROSS MODALITY COVERAGE, IOT ARCHITECTURE, AND FORMAL DIGITAL TWIN CRITERIA

| Paper | Modalities | IoT Architecture | State Space | Transition Model | Digital Twin Criteria Satisfied |
|---|---|---|---|---|---|
| Herborn et al. [9] | Acoustic | No | No | No | No |
| Manikandan & Neethirajan [5] | Acoustic | No | No | No | No |
| Li et al. [4] | Video + Acoustic (review) | No | No | No | No |
| Dawkins et al. [11] | Video | No | No | No | No |
| Neethirajan & Kemp [18] | Conceptual Review | Conceptual only | No | No | No |
| Essien & Neethirajan [14] | Thermal + Acoustic + Video + Environmental (Review) | No | No | No | No |
| Garro & Sorrenti [8] | Industrial sensors | Yes (ISO 23247) | Yes | Yes | Partial |
| HenTwin (This work) | Thermal +Acoustic +Video+ | Yes (5-layer IoT) | Yes | Yes | Yes |

Environm
ental

HenTwin represents the application of a formally estimated discrete-time state-space model to multimodal poultry state data, integrating thermal, acoustic, behavioural, and environmental components. This positions HenTwin at the intersection of precision livestock farming, digital twin engineering, and biological state-space modelling, contributing a methodology that is informed by welfare science and formally aligned with the system-theoretic requirements of a digital twin.

## III. Experimental setup

### A. Animals and Housing

The present study was conducted at the Atlantic Poultry Research Centre of the Dalhousie Agricultural Campus in Truro, Nova Scotia, Canada, with all procedures approved by the institutional Animal Care and Use Committee under Protocol No. 2025-012. A cohort of 150 Lohmann LSL-Lite chicks was housed across 5 controlled environmental rooms at a stocking density of 30 birds per room. Room temperature was managed under a gradual brooding regime, declining from approximately 35°C at placement to 21°C by the end of week four, after which conditions were stabilized at the rearing setpoint and maintained for the remainder of the trial. Birds were exposed to a 16:8 hour light-to-dark photoperiod and provided ad libitum access to feed and water throughout. Multimodal data acquisition proceeded continuously from June to November 2025, spanning 25-consecutive weeks. Although instrumentation was installed in all five rooms, the analyses reported in this study concentrate on Rooms 1 and 5 as the primary study units, since these were the only rooms in which complete acoustic and environmental records were maintained across the full 25-week monitoring window.

### B. System Architecture Overview

HenTwin's architecture is shaped by three deployment constraints. First, raw video and audio streams generate data volumes impractical for continuous cloud transmission over barn-grade connectivity. Edge-level feature extraction reduces each modality to a single scalar per week, making upload feasible without dedicated high-bandwidth infrastructure. In practical terms, continuous video and audio acquisition produces on the order of tens of gigabytes of raw data per room per week, whereas the edge pipelines upload only a handful of scalar features per modality per week, reducing cloud communication to approximately tens of bytes. Second, flock dynamics vary across rooms and facilities and a shared transition structure with lightweight local calibration of environmental input sensitivity enables deployment across multiple barns without repeating the full longitudinal study. Third, farm operators need actionable outputs rather than raw sensor feeds the architecture culminates in a decision-support layer providing early alerts about deviations in multimodal state and what-if simulation, allowing operators to ask questions such as "what happens to flock acoustic state if THI remains above 72 for the next three weeks?" and receive a model-projected answer. These three goals of edge compression, transferable model structure, and operator-facing simulation directly motivate the layered design described below.

HenTwin is organized as a five-layer Internet-of-Things architecture that transforms raw multimodal barn signals into a digital-twin representation of the flock's multimodal biological state (Fig. 1). The layered design separates data acquisition, on-site feature extraction, state-space modelling, the digital-twin abstraction, and operator-facing decision support, and makes explicit which elements are implemented and validated in the present study and which define the proposed deployment pathway.

At the base, the sensing layer acquires four complementary modalities – one environmental channel (microclimate) and three biological-response channels (thermal, acoustic, and visual) – providing the heterogeneous signal base required for a genuinely multimodal state representation. The edge processing layer reduces each raw stream to a compact, biologically interpretable feature through a modality-specific pipeline, yielding the temperature–humidity index, body surface temperature, the acoustic key performance indicators, and a motion baseline. The modelling layer assembles these features into the four-dimensional state vector and estimates the discrete-time state-space model “$S(t+1) = A\cdot S(t) + c + B\cdot u(t) + w(t)$”, in which the temperature–humidity index enters as an exogenous input rather than a state component and a parameter store retains the site-specific coefficients. The digital twin layer provides the virtual representation of the flock, supporting forward perturbation simulation and change-point detection over the modelled state, while the decision-support layer is the proposed operator-facing tier. A feedback path from the modelling layer to the edge layer realizes a site-specific calibration loop, by which a shared model structure is adapted to a new deployment context using a short local observation period rather than full re-estimation.

The modelling and perturbation components of the framework are implemented and validated from data in the present study, whereas a live, continuously updating state mirror and the operator-facing decision-support tier define the proposed deployment pathway. The subsections that follow describe each layer in turn: the sensing suite (Section III.C), the per-modality feature extraction performed at the edge (Section IV), and the state-space model and digital-twin formulation (Section V).

### C. Sensor Suite and Data Acquisition

Four sensing modalities operated concurrently to characterize the environmental, physiological, acoustic, and behavioural dimensions of flock state. The following subsections describe the sensing modalities, sampling protocols, and processing pipelines applied to each modality.

1) *Environmental monitoring:* Two complementary environment monitoring systems were deployed – a primary system providing the core environmental inputs to the HenTwin model, and a secondary system providing contextual air-quality data. The principal system consisted of wall-mounted DHT22 digital sensors calibrated to within ±2 percent accuracy. These units logged ambient temperature and relative humidity at two fixed time points each day, 08:00 and 16:00 hours, in each controlled room. The secondary system extended this primary monitoring to a wider air-quality profile through daily measurement of carbon dioxide concentration, volatile organic compound levels, and particulate matter indices using portable multi-sensor arrays (Model Air-Guardian AQ, Simbow Inc., Shenzhen, Guangdong, China). The supplementary air-quality variables were not used as inputs to the HenTwin model in the present study. They were recorded for descriptive contextualization of room conditions and to inform future extension of framework. In combination, the two monitoring streams generated the continuous environmental record required to interpret the physiological and behavioural responses observed across the rearing period. Fig. 2 illustrates the in-room deployment of the environmental monitoring setup, with the multi-parameter air-quality display unit installed on a fixed wall-mounted configuration to maintain consistent measurement geometry across the study period.

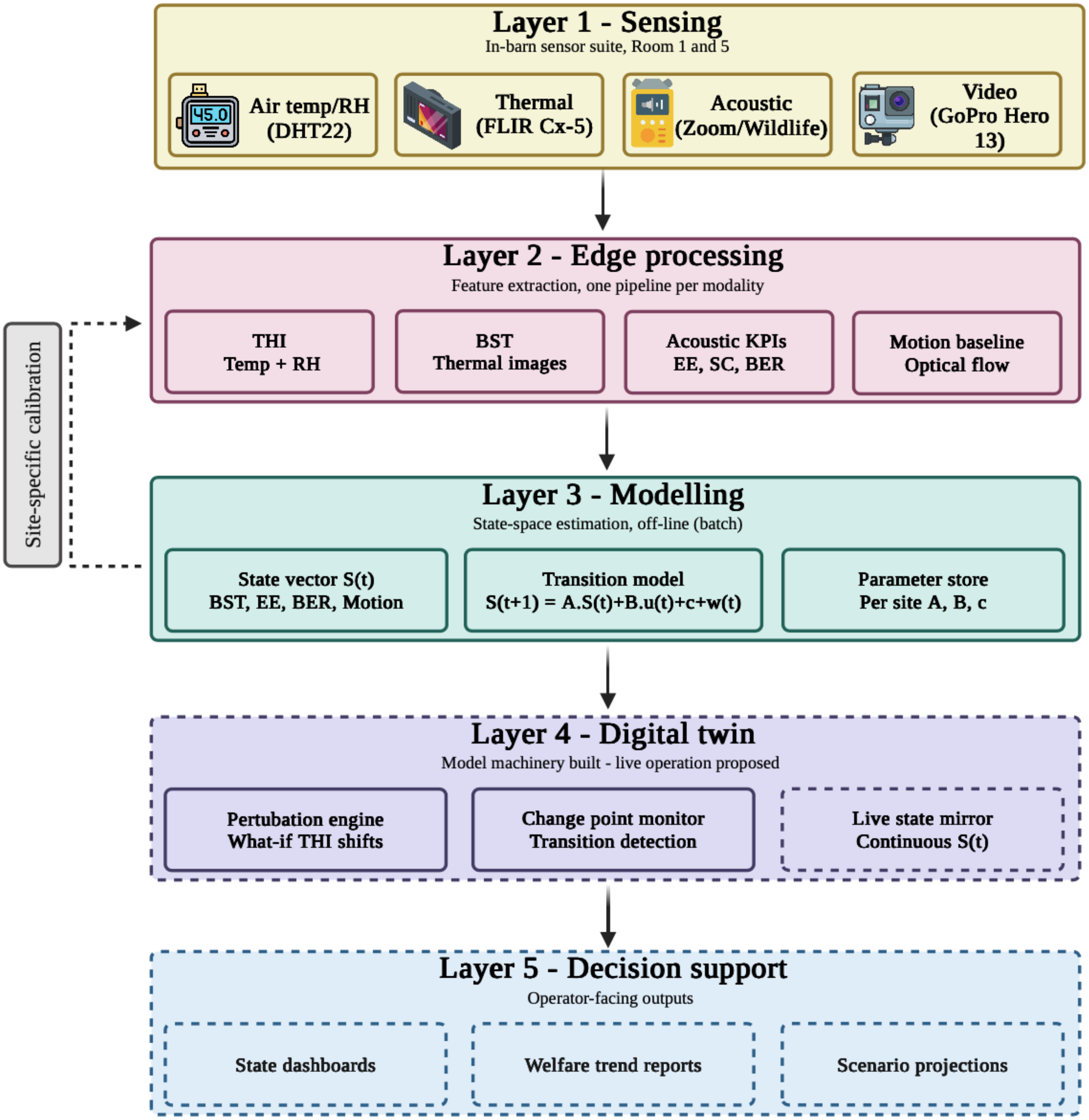

Fig. 1. HenTwin five-layer IoT architecture. Multimodal barn signals are acquired by the sensing layer (Layer 1) across four modalities – microclimate (DHT22), thermal (FLIR Cx-5), acoustic (Zoom/Wildlife recorder), and visual (GoPro Hero 13). The edge processing layer (Layer 2) reduces each modality to a compact feature through a dedicated pipeline: the temperature–humidity index (THI) from microclimate, body surface temperature (BST) from thermal frames, acoustic key performance indicators (energy entropy (EE), spectral centroid (SC), band energy ratio (BER)) from audio, and a motion baseline from optical flow. The modelling layer (Layer 3) assembles these features into the state vector $S(t) = [BST, EE, BER, motion]^T$ and estimates the discrete-time state-space model $S(t+1) = A{\cdot}S(t) + c + B{\cdot}u(t) + w(t)$, where THI enters as the exogenous input $u(t)$ and $w(t)$ is a zero-mean Gaussian noise term. Per-site parameters (A, B, c) are held in a parameter store. The digital twin layer (Layer 4) hosts the perturbation engine and change-point monitor, and the decision-support layer (Layer 5) exposes operator-facing outputs. A dashed feedback loop returns site-specific calibration from the modelling layer to the edge layer, realizing the two-tier deployment strategy. Solid components are implemented and validated from data in this study whereas dashed components define the proposed deployment pathway.

From the temperature and humidity measurements, the Temperature-Humidity Index (THI) was computed using the layer-specific formulation of Zulovich and DeShazer [22]:

$$THI_{layers} = 0.6 \times T_{db} + 0.4 \times T_{wb} \quad (1)$$

where $T_{db}$ represents the dry-bulb temperature in degrees Celsius and $T_{wb}$ represents the wet-bulb temperature in degrees Celsius, with the resulting index expressed in degrees Celsius. The morning and afternoon THI values were averaged to derive a single daily estimate, and these daily values were then averaged across the seven days of each study week to yield a weekly THI trajectory specific to each instrumented room. This procedure was applied independently to Rooms 1 and 5, producing two parallel room-resolved weekly profiles of environmental thermal load across the entire 25-week study.

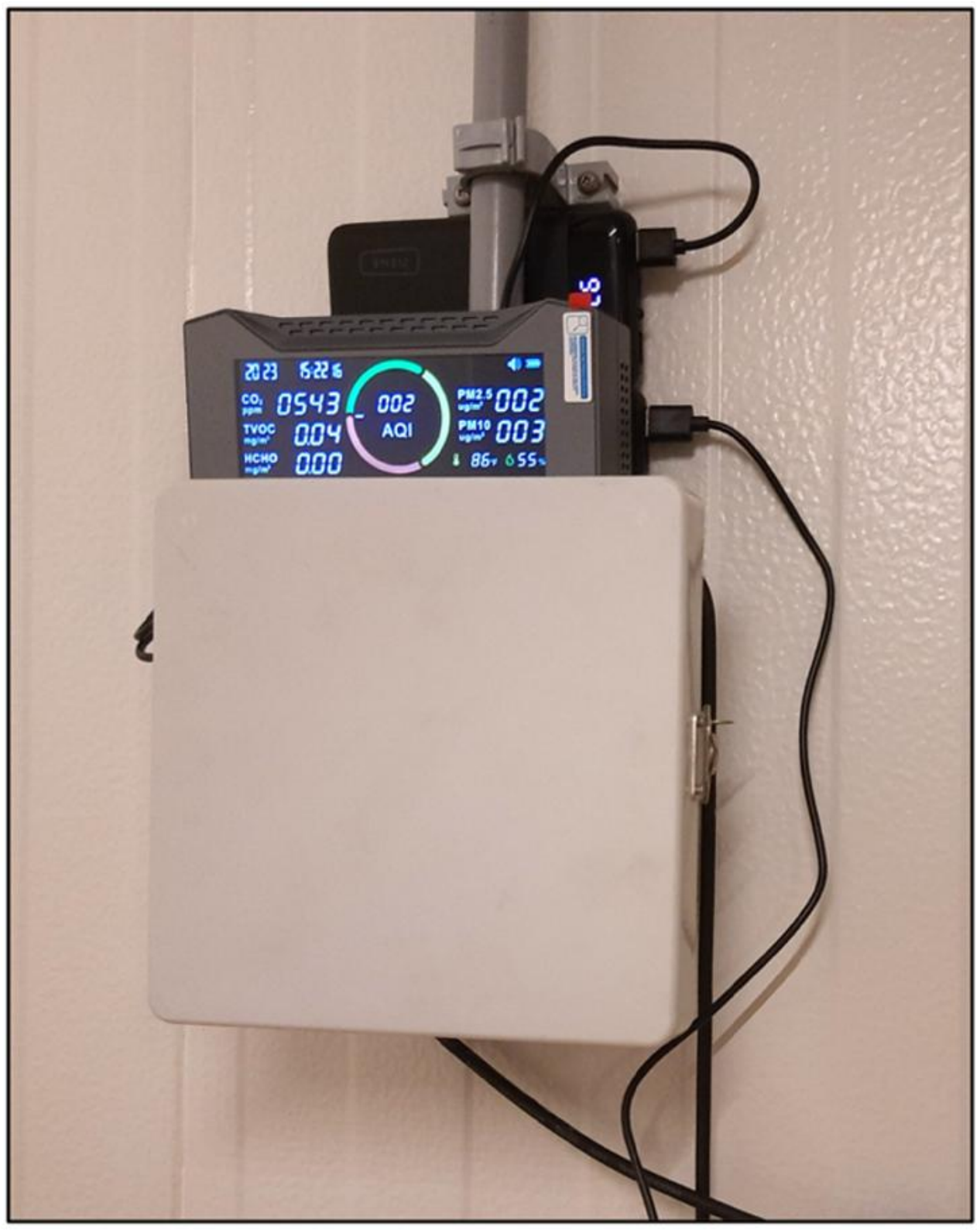

Fig. 2. In-room environmental monitoring setup featuring a wall-mounted Simbow Air-Guardian AQ multi-parameter unit providing real-time readings of $CO_2$, TVOC, HCHO, PM2.5, PM10, temperature, and humidity, operating alongside DHT22 digital sensors for primary temperature and relative humidity logging. Fixed wall placement and battery-supported operation ensured continuous, temporally consistent records across the 25-week study.

2) *Thermal imaging:* Body surface temperature was measured using a FLIR Cx-5 radiometric thermal imaging camera (Model Cx-5, FLIR Systems Inc., Wilsonville, OR, USA), with emissivity configured to 0.95 and the lens positioned at a standardised working distance of approximately one metre from each photographed bird. Imaging was performed at 48-hour intervals throughout the study, with each session generating 6 to 8 thermal images per room. Aggregated across the five controlled rooms, this acquisition cadence produced approximately 35 to 40 images each imaging day and between 140 and 160 images per week on a farm-wide basis. Fig. 3 presents the thermal imaging workflow, with panel (a) illustrating the handheld in-room acquisition procedure and panel (b) showing a representative output image used for subsequent regional annotation and body surface temperature extraction. From this larger image pool, a curated subset of 25 images per week was retained for manual annotation and downstream analysis, selected on the basis of image sharpness, clear visibility of head and foot regions, and the absence of motion blur or partial occlusion. Across the full 25-week monitoring period, this selection protocol yielded a final annotated dataset of 625 thermal images. Each retained image was uploaded to the FLIR Ignite cloud platform, where head and foot regions of interest were delineated manually to extract maximum, minimum, and mean Body Surface Temperature (BST) values.

Because thermal images were not labelled by room of origin during acquisition, the resulting BST series reflects a farm-wide weekly average rather than room-resolved measurements. This is acknowledged as a methodological limitation and is addressed explicitly in the discussion. To maintain consistency with the room-specific treatment of the other modalities, the same farm-wide BST series was used as the thermal input for both Rooms 1 and 5 throughout the downstream analyses, including the state-space modelling and perturbation simulation. Future deployments of HenTwin will adopt room-tagged thermal acquisition to remove this limitation.

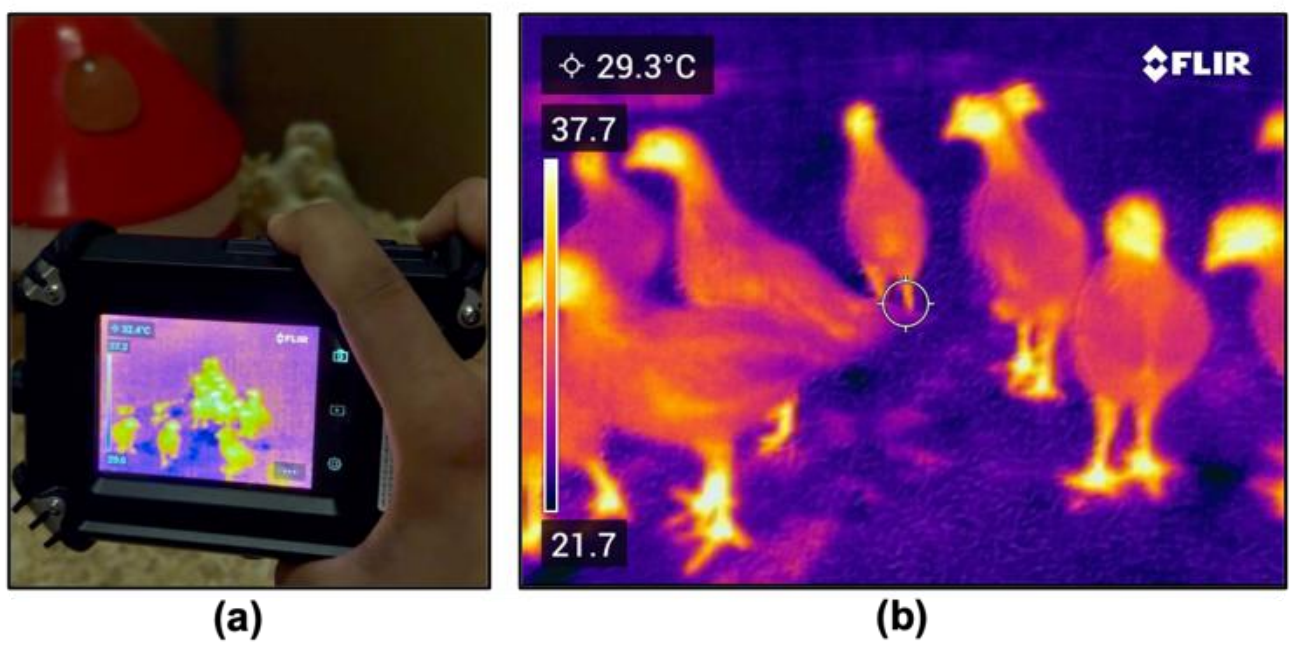

Fig. 3. Thermal imaging workflow for Body Surface Temperature (BST) acquisition. (a) Handheld in-room capture of thermal images using a FLIR Cx-5 radiometric camera positioned at a standardised working distance of approximately one metre from the birds. (b) Representative radiometric thermal image of laying hens, showing surface temperature distribution across the flock with a measured spot value of 28.7°C and a scene range of 20.5 to 36.1°C. Head and foot regions were subsequently delineated manually on each retained image to extract maximum, minimum, and mean BST values.

3) *Acoustic recording:* Acoustic monitoring was implemented through a four-unit recording array selected to balance portability, sensitivity, and unattended long-duration operation in four of the five rooms. Two Zoom H4n Pro recorders (Model H4n Pro, Zoom North America, Hauppauge, NY, USA) were each fitted with an external RØDE lavalier microphone (Model NTG-2, RØDE Microphones LLC (USA), Long Beach, CA, USA), and a third channel was provided by a Zoom F6 professional field recorder (Model F6, Zoom North America, Hauppauge, NY, USA) connected to a further external RØDE microphone. Furthermore, continuous unattended recording in another room was provided by a Wildlife Acoustics Song Meter SM4 autonomous recorder (Model SM4, Wildlife Acoustics Inc., Maynard, MA, USA). The Zoom recorders were powered by portable rechargeable battery packs, while the Song Meter operated from four rechargeable D-size NiMH cells. This combination of power configurations enabled the array to sustain uninterrupted recording over extended acquisition windows. Recording proceeded continuously throughout the daily photoperiod for the entire 25-week study, producing near-continuous whole-room audio coverage that captured both routine vocal activity and event-driven responses. Fig. 4 shows the acoustic recording array deployed in the study rooms, with each of the four recording units installed in a separate environmental room to provide continuous whole-room audio coverage across the 25-week monitoring period.

All microphones were installed at consistent heights of 1.5 to 2.0 meters above the litter surface, supporting the capture of whole-room ambient sound under uniform gain settings across rooms and across the full study period. Audio was sampled at 44.1 kHz with 16-bit resolution and stored in uncompressed WAV format to preserve signal fidelity for downstream spectral analysis and to support archival access for future re-analysis. Although recording was continuous, quantitative feature extraction was performed on a curated subset of the acoustic stream rather than the full archive. Specifically, between 10 and 12 one-minute clips were sampled per week from the peak morning vocal activity window of 06:00 to 09:00 hours on days 1, 4, and 7 of each study week. This sampling design preserved a clear separation between the continuous archival recording and the targeted subset used for feature extraction, ensuring that the analyses were based on temporally consistent and reproducible audio segments while retaining the underlying recordings for future re-investigation.

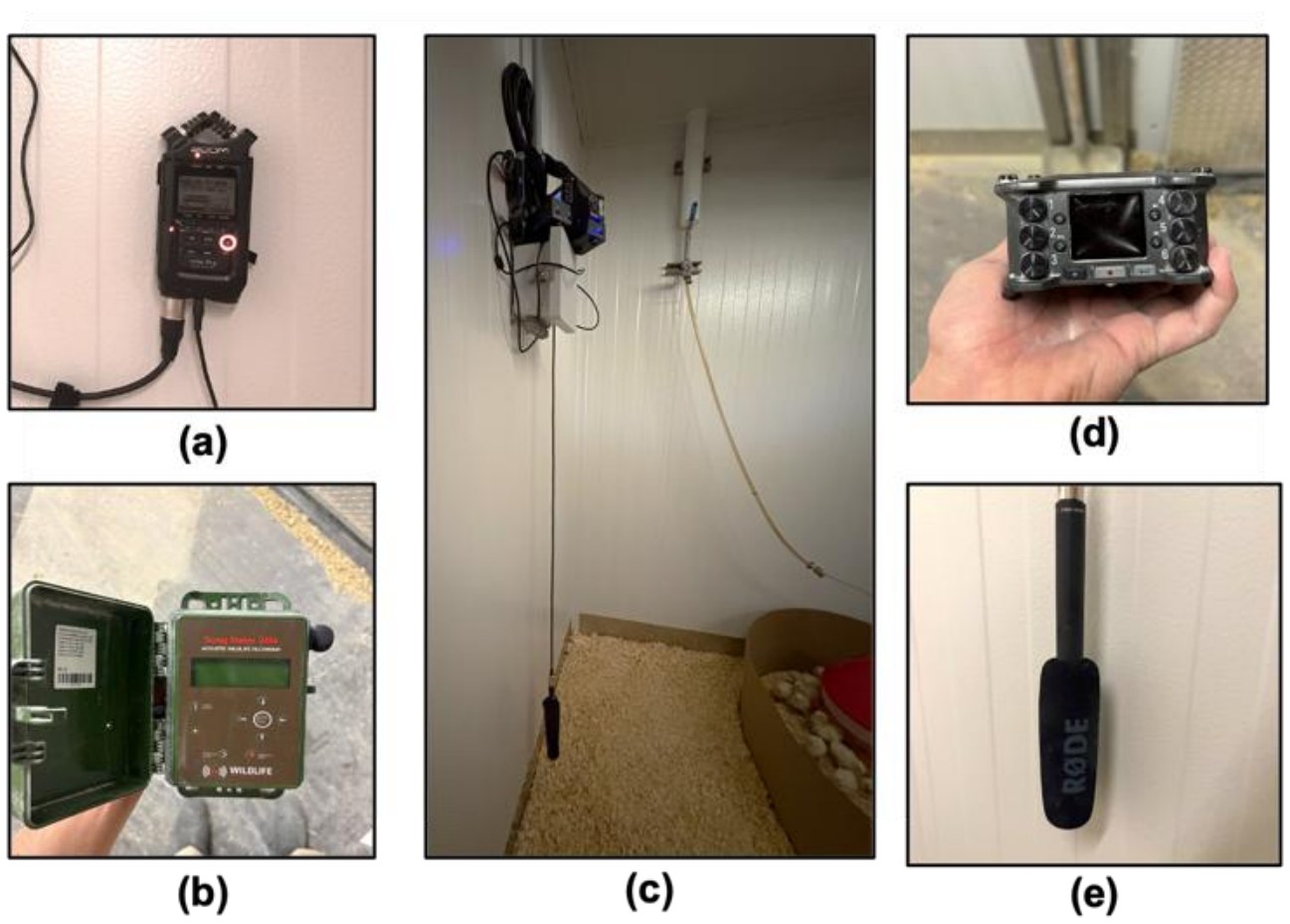

Fig. 4. HenTwin acoustic recording array deployed across the study rooms. (a) Wall-mounted Zoom H4n Pro handheld recorder capturing continuous whole-room audio. (b) Wildlife Acoustics Song Meter SM4 autonomous recorder used for unattended long-duration recording. (c) In-room deployment of the Zoom F6 professional field recorder with battery support for extended operation (d) Front view of the Zoom F6 showing multi-channel input controls and integrated display. (e) External RØDE NTG-2 lavalier microphone connected to the Zoom recorders for high-fidelity capture of flock vocalizations.

4) *Video acquisition:* Behavioural video data were captured using overhead GoPro Hero 13 cameras (Model HERO13 Black, GoPro Inc., San Mateo, CA, USA) mounted in fixed positions at roof within each room. The cameras recorded continuously at 1080p resolution and 30 frames per second. Video collection commenced at week 5 rather than at the outset of the study, since the cameras could not be operated reliably during weeks 1 to 4. The elevated brooding temperatures required to support early chick development exceeded the operational thermal tolerance of the cameras, causing intermittent shutdowns during that initial period. From week 5 onward, the cameras were maintained on continuous AC power, enabling uninterrupted recording through to week 25 and yielding 21 consecutive weeks of standardised video data with consistent camera geometry. Fixed mounting positions preserved spatial consistency across all recordings and supported direct comparison of behavioural dynamics across the rearing trajectory. Fig. 5 illustrates the video acquisition setup, with panel (a) showing the ceiling-mounted GoPro Hero 13 camera positioned to capture overhead views of the flock and panel (b) showing the direct AC power connection from the room power outlet to the camera, enabling continuous uninterrupted recording across the study period.

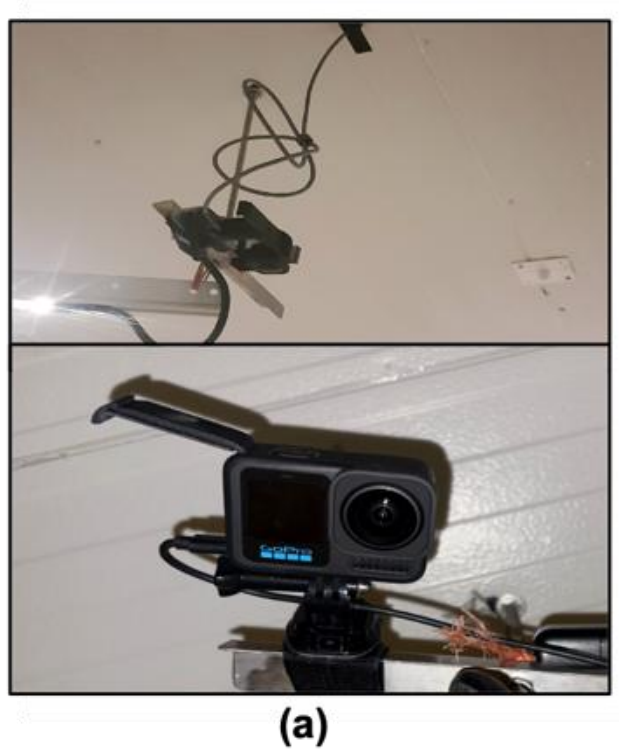

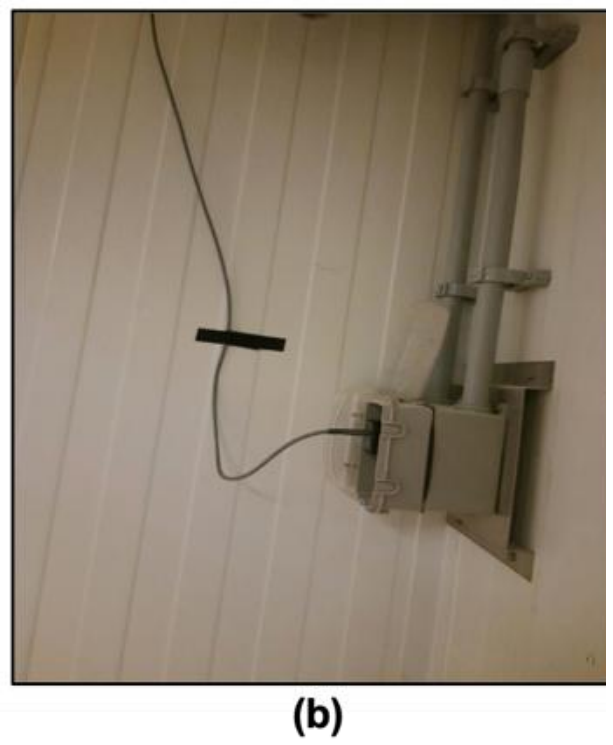

(a) (b)

Fig. 5. Video acquisition setup for behavioural monitoring in Room 1. (a) Ceiling-mounted GoPro Hero 13 camera installed in a fixed overhead position to provide a consistent top-down view of the flock at 1080p resolution and 30 frames per second. (b) Direct AC power connection from the room power outlet to the camera, enabling continuous uninterrupted recording from week 5 through week 25 of the study.

## IV. Data Processing and Feature Extraction

### A. *Acoustic Feature Extraction*

Each one-minute audio clip was processed using the librosa library in Python 3.9 to compute a panel of spectral and energy-based features. Three primary acoustic Key Performance Indicators (KPIs) were computed as candidate descriptors of the flock vocal state. Energy Entropy (EE) quantifies the disorder and complexity of the spectral energy distribution and is sensitive to changes in vocal complexity associated with thermal and physiological stress. Spectral Centroid (SC) captures the centre of mass of the audio spectrum and is sensitive to shifts in vocal frequency content. Band Energy Ratio (BER) in the 2 to 5 kHz frequency band quantifies the proportion of total energy concentrated in the band associated with high-frequency distress calls in laying hens.

Three additional acoustic features – zero-crossing rate, root mean square energy, and spectral flatness – were initially computed during exploratory analysis. These features were excluded from the primary HenTwin state vector for two reasons. First, they did not exhibit consistent relationships with the established welfare indicators in the present dataset. Second, they lack sufficient precedent in the peer-reviewed poultry welfare literature to justify their inclusion in a multimodal framework that prioritizes biologically interpretable indicators. The acoustic feature set therefore comprises EE, SC, and BER as biologically interpretable descriptors, each computed at the weekly aggregation level by averaging across all clips within each week. Of these, EE and BER are carried into the state vector while SC is excluded during estimation, with the rationale detailed in Section V.A.

### B. *Video-Based Behavioural Feature Extraction*

1) *Video selection and preprocessing:* Video data were processed for Room 1 across weeks 5 through 25, corresponding to the period during which reliable camera operation was achievable in the post-brooding environment. Three 15-minute video segments were selected per week, recorded on days 1, 4, and 7 and centred on routine caretaker entry events. Each segment was partitioned into three behavioural conditions – a 7-minute pre-entry baseline period, a 1 to 2-minute active caretaker presence period, and a 7-minute post-entry recovery period. Each condition was further divided into non-overlapping 30-second analysis windows to enable localized temporal characterization of motion patterns while minimizing the influence of frame-level noise.
   All frames were preprocessed prior to motion estimation through a consistent pipeline. Frames were converted to grayscale to reduce computational overhead and mitigate variation arising from lighting changes. Each grayscale frame was resized to a uniform spatial resolution of 640 × 360 pixels. Where appropriate, frames were spatially cropped to a fixed region of interest corresponding to the bird-occupied floor area, excluding static structural elements such as enclosure boundaries, feeders, and lighting fixtures. A mild Gaussian smoothing operation was applied to suppress high-frequency noise while preserving motion-relevant spatial gradients.

2) *Dense optical flow estimation:* Motion between consecutive frames was quantified using dense optical flow based on the Dense Inverse Search (DIS) algorithm implemented in OpenCV 4.5. This method was selected because it provides reliable pixel-wise motion estimates at substantially lower computational cost than alternative dense methods such as Farnebäck flow, making it suitable for large-scale analysis of long video recordings. For each frame pair, the algorithm produced a two-dimensional motion vector field:
   $$V(x, y, t) = [\, u(x, y, t),\ v(x, y, t)] \tag{2}$$
   where u and v denote the horizontal and vertical components of pixel-wise displacement between frames. To reduce computational cost without sacrificing temporal resolution for behavioural characterization, optical flow was computed on alternating frame pairs, corresponding to an effective processing rate of 15 frame pairs per second. The scalar motion intensity at each pixel was derived from the flow vector field as:
   $$M(x, y, t) = \sqrt{u(x, y, t)^2 + v(x, y, t)^2} \tag{3}$$
   which provides a directionally invariant representation of instantaneous motion intensity across the spatial domain.

3) *Behavioural feature derivation:* Within each 30-second analysis window, the spatial distribution of motion magnitude was summarized into a set of complementary descriptors capturing different aspects of collective flock movement. These included mean flow magnitude as a measure of overall flock activity level, the standard deviation of flow magnitude as a measure of spatial heterogeneity of movement, the 95th percentile of flow magnitude as a measure of burst-motion intensity, motion entropy computed from the normalized histogram of motion magnitudes as a measure of behavioural disorder, and the active pixel fraction defined as the proportion of pixels exceeding a motion threshold of 0.3 units as a measure of the spatial extent of movement. Window-level features were aggregated within each behavioural condition (before, during, after) and then across all clips within each study week.

Three derived behavioural indices were further computed to characterize the dynamics of flock response to caretaker entry events. The Entry Reactivity Index was defined as the difference between mean flow during caretaker presence and mean flow during the baseline period, capturing the magnitude of acute behavioural response. The Recovery Index was defined as the difference between mean flow after caretaker departure and mean flow during entry, characterizing the speed and completeness of post-disturbance recovery. The Disturbance Amplification Index was defined as the ratio of 95th percentile flow during entry to median flow during the baseline period, capturing the relative intensity of behavioural disturbance events. The mean flow baseline metric, representing background flock activity in the absence of disturbance events, was selected as the primary motion component of the HenTwin state vector based on a sensitivity analysis described in Section V.D.

## V. HenTwin State-Space Formulation

### A. Biological State Vector and Environmental Input

The HenTwin digital twin represents the flock biological state at discrete weekly intervals through a multimodal state vector. The state at time step *t* is defined as:

$$S(t) = [BST(t), EE(t), BER(t), Motion(t)]^T \tag{4}$$

where BST(t) denotes the weekly mean body surface temperature in degrees Celsius, EE(t) denotes the weekly mean Energy Entropy in nats, BER(t) denotes the weekly mean Band Energy Ratio in the 2 to 5 kHz band expressed as a percentage, and Motion(t) denotes the mean optical flow baseline for the week. The Temperature-Humidity Index is explicitly treated as an external environmental input rather than a component of the biological state vector:

$$u(t) = THI(t) - E[THI] \tag{5}$$

where E[THI] denotes the long-run average THI computed across the full 25-week observation period. This separation between the biological state and the environmental input is a deliberate architectural decision. Including THI inside the state vector while simultaneously using it as an input to drive that vector would create an artificial feedback loop in which the environmental forcing variable would propagate through the state equations and amplify itself across successive time steps. By treating THI as a strictly exogenous input that drives the biological system without being driven by it, HenTwin captures the correct causal direction – environmental conditions influence flock biology, not the reverse – and acquires genuine intervention capability whereby alternative THI scenarios can be specified to project the corresponding biological state response.

Spectral Centroid was initially included in exploratory state vector formulations but was subsequently excluded from the transition model on two grounds. First, SC did not exhibit a statistically significant linear coupling to THI at the weekly aggregation level ($p = 0.225$). Second, SC values fall in a numerical range of approximately 300 to 3000 Hz, roughly two orders of magnitude larger than the other state variables. This scale mismatch introduces numerical instability when SC is included in a multivariate state transition without careful standardisation, particularly given the limited sample size available for parameter estimation. The four-component state vector defined above therefore represents the parsimonious set of biologically interpretable variables with stable scaling and significant environmental coupling.

### B. Discrete-Time State Transition Model

The temporal evolution of the biological state vector is modelled as a discrete-time linear dynamical system:

$$S(t+1) = A \cdot S(t) + c + B \cdot u(t) + w(t) \tag{6}$$

where A is the state transition matrix capturing the intrinsic developmental dynamics of each state component, c is a vector of regression intercepts, B is the input coupling vector mapping environmental THI deviation to changes in each biological state variable, u(t) is the scalar THI deviation defined in the preceding section, and w(t) is a zero-mean Gaussian noise term absorbing biological variability not explained by the model.

Given the limited available sample size of 20 weekly state transitions in the training period, the transition matrix A was constrained to diagonal form. This constraint is methodologically necessary rather than merely convenient. Estimating a full $4 \times 4$ cross-variable transition matrix requires the identification of 16 parameters from 20 observations, which produces an underdetermined estimation problem and severe overfitting. Empirical verification confirmed that an unrestricted A matrix produced simulated state trajectories that diverged to physically impossible values, including negative Energy Entropy. Constraining A to its diagonal form reduces the number of free parameters in the transition matrix to four – one persistence coefficient per state variable – yielding a 5:1 ratio of observations to parameters that supports stable estimation. The diagonal model therefore characterizes each state variable as a first-order autoregressive process driven by its own lagged value and the external THI input, with cross-variable interactions absorbed into the residual error term rather than explicitly modelled.

The diagonal entries of A and the intercept vector c were estimated by ordinary least squares regression of each variable at time t+1 against its own value at time t. The estimated persistence coefficients were 0.668 for BST, 0.831 for EE, 0.543 for BER, and 0.383 for Motion baseline. All four values are strictly less than unity, confirming that the transition model is stable and that simulated trajectories converge rather than diverge over time. The largest eigenvalue of the diagonal A matrix has magnitude 0.831, well within the stability region of the unit circle. Energy Entropy exhibits the highest persistence, indicating that vocal complexity changes slowly across consecutive weeks under average environmental conditions, while motion baseline exhibits the lowest persistence, indicating that flock activity exhibits more rapid week-to-week reversion to its mean level.

### C. Steady-State Characterization of the Biological State

The steady-state value of each biological state variable, defined as the fixed point of the transition equation under zero environmental input, was derived analytically. Setting S(t+1) = S(t) = S* and u(t) = 0 in the transition equation and rearranging yields:

$$S^{*} = c \,/\, \left(1 - A_{diag}\right) \tag{7}$$

where the division is performed elementwise across the four state components. Substituting the estimated parameters produces steady-state values of 24.94°C for BST, 7.56 nats for EE, 4.44% for BER, and 0.582 for Motion baseline. These analytical predictions correspond closely to the observed weekly means across the study period (25.03°C, 7.86 nats, 7.43%, and 0.572 respectively), with the residual differences attributable to the elevated values observed during the early brooding phase that raise the empirical mean slightly above the long-run steady state. The agreement between the analytical steady-state values and the empirical means provides an internal consistency check on the estimated transition model. The steady-state vector S* also serves as the principled initial condition for the forward simulation analyses described in the following section, as initialization at S* removes startup transients associated with arbitrary observed starting points and isolates the sustained effect of environmental perturbations.

### D. Estimation of Environmental Input Coupling

The input coupling vector B quantifies the sensitivity of each biological state variable to deviations of THI from its long-run average. Each coefficient was estimated by ordinary least squares regression of the biological variable on the contemporaneous THI deviation:

$$x_{i(t)} = B_i \cdot u(t) + k_i + \varepsilon(t) \tag{8}$$

where $x_i$ denotes the i-th biological state variable, $k_i$ is a regression intercept, and ε(t) is the residual. This direct estimation isolates the equilibrium association of each channel with THI. Jointly estimating the coupling and persistence terms on the 20-observation series yields markedly wider coupling intervals (Section VII.G), so the direct estimate is preferred for the structural and perturbation analyses. Estimated this way, each coefficient represents the contemporaneous, equilibrium-level association between THI deviation and the state variable within the modelling window. This is the quantity used to characterize the steady-state and perturbation behaviour in Sections V.E and VII.D. The direct procedure produces interpretable point estimates and confidence intervals for B without the compounding estimation error associated with multi-stage methods on short time series.

Only coupling coefficients with associated p-values below the 0.05 threshold were retained in the simulation model; coefficients failing this significance criterion were set to zero to avoid the propagation of statistical noise into the predictive analyses. Spectral Centroid was excluded from the transition model on this basis ($p = 0.225$), confirming the prior decision based on numerical scaling considerations. The retained B coefficients were +0.300°C per THI unit for BST ($R^2 = 0.462$, $p < 0.001$), +0.271 nats per THI unit for EE ($R^2 = 0.725$, $p < 0.001$), +3.334 percentage points per THI unit for BER ($R^2 = 0.311$, $p = 0.009$), and −0.037 units per THI unit for Motion baseline ($R^2 = 0.327$, $p = 0.007$). The signs of these coefficients are biologically coherent: higher thermal load elevates body surface temperature, vocal complexity, and high-frequency distress call energy, while simultaneously suppressing baseline flock activity through behavioural thermoregulation.

A sensitivity analysis was conducted to verify that mean flow baseline rather than alternative motion descriptors was the appropriate choice for the motion component of the state vector. The Entry Reactivity Index was tested as an alternative motion variable using the same direct estimation procedure. The resulting B coefficient for reactivity was +0.017 per THI unit with $R^2$ of 0.096 and $p = 0.172$, which did not satisfy the significance threshold for inclusion in the transition model. This result confirms that motion reactivity, defined specifically in the context of caretaker entry events, captures event-driven behavioural responses that are not reliably coupled to ambient thermal conditions at the weekly aggregation level, and that mean flow baseline provides a more stable and environmentally responsive state component.

The robustness of the input coupling coefficients was assessed using a bias-corrected and accelerated (BCa) bootstrap with 2,000 resamples of the training weeks. The direct single-variable estimates excluded zero for all four channels and were therefore retained as the reported couplings for the structural and perturbation analyses. The same couplings estimated jointly with the persistence terms — the configuration used as a one-step forecasting baseline in Section VII.G – carried substantially wider intervals on the 20-observation series, reflecting collinearity between each channel's previous state and the contemporaneous THI input rather than an absence of coupling. The direct estimates are consequently the appropriate quantity for the equilibrium and perturbation analyses, while the jointly estimated form is reported only in the forecasting comparison.

### E. Forward Simulation and Perturbation Analysis

The estimated state-space model supports forward simulation of the biological state trajectory under both observed and hypothetical environmental scenarios. To assess the intervention-aware capability of the HenTwin framework, a perturbation analysis was conducted in which a sustained environmental shock was applied to the THI input and the corresponding biological state response was simulated forward through time. The unperturbed simulation evolves according to the transition equation with the empirically observed THI deviation series, while the perturbed simulation introduces an additive shock δ to the input:

$$S_{baseline(t+1)} = A \cdot S_{baseline(t)} + c + B \cdot u(t) \tag{9}$$

$$S_{perturbed(t+1)} = A \cdot S_{perturbed(t)} + c + B \cdot \left(u(t) + \delta\right) \tag{10}$$

Both simulations were initialized at the analytically derived steady state S* described in Section V.C, ensuring that any divergence between the two trajectories is attributable solely to the applied perturbation rather than to differences in initial conditions. The predicted impact of the perturbation at each time step is defined as the difference between the perturbed and baseline state trajectories:

$$\Delta S(t) = S_{perturbed(t)} - S_{baseline(t)} \tag{11}$$

As the simulation progresses forward in time, ΔS(t) converges to a stable long-run impact ΔS*, the sustained equilibrium consequence of the persistent environmental shift. For this linear model the long-run impact equals the product of the equilibrium input coupling and the applied shift, $B \cdot \delta$, computed elementwise. This equilibrium impact is the primary reported outcome of the perturbation analysis.

A perturbation magnitude of $\delta = +2.0$ THI units was selected for the primary analysis on empirical grounds. The observed weekly THI deviations across the 25-week study period spanned the range from −2.26 to +2.26 units, indicating that a +2.0 THI shock represents the upper bound of natural weekly variation originally observed in this dataset. This choice grounds the simulation in empirically realistic conditions rather than hypothetical extremes and produces results that are directly interpretable in terms of conditions the flock could plausibly encounter under normal operational variability. The perturbation was applied from week 5 onward, corresponding to the point at which the graduated brooding temperature regime had stabilized, the full four-modality state vector became available, and routine video monitoring commenced. Band Energy Ratio predictions were constrained to non-negative values throughout the simulation, reflecting the physical interpretation of BER as a percentage that cannot fall below zero. Because band energy ratio is non-stationary over the study period (Section VII.H), its perturbation response is reported for completeness but is not interpreted as a robust quantitative projection.

### *F. Digital Twin Functions*

The estimated state-space model instantiates three capabilities that distinguish HenTwin as a digital twin rather than a monitoring pipeline.

The first capability is forward simulation under counterfactual environmental scenarios. Given any hypothetical THI trajectory, the transition equation propagates the biological state forward in time, producing projected trajectories for BST, EE, BER, and Motion without requiring new sensor data. Concrete operator questions – for example, *if ventilation failure raises THI by 2 units above the stable baseline for four weeks, what happens to flock acoustic entropy?* – are answered directly by the perturbation analysis. The perturbation analysis answers this as "*EE is projected to rise by 0.54 nats at steady state*". This provides a quantitative basis for proactive intervention rather than reactive response.

The second capability is data-driven detection of developmental state transitions. Applying Pettitt change-point analysis to the modelled state trajectory, HenTwin identifies the weeks at which the flock's biological state shifts in mean level, without manual threshold definition. The near-simultaneous detection of a coordinated transition across three independently measured channels within weeks 12–14 (Section VII.E) demonstrates that the framework localizes system-level developmental changes from the multimodal state itself, rather than from any single sensor.

The third capability is cross-room transferability through parameter sharing. The two-tier architecture with the shared transition matrix A, room-specific input coupling B, means a new barn requires only a short local calibration period rather than a full longitudinal study. This reduces deployment cost and time while preserving predictive accuracy, a practical requirement for scaling IoT-based welfare monitoring across commercial facilities.

Together these three capabilities align with the principles underlying the ISO 23247 digital twin framework as formalized in [8].

## VI. Statistical and Validation Analyses

### *A. Cross-Modal Correlation Analysis*

Cross-modal coupling between state vector components and the environmental input was quantified using Spearman rank correlation. Spearman correlation was selected in preference to Pearson correlation because biological time series at weekly aggregation are typically non-normally distributed and may contain non-linear monotonic relationships that are better captured by rank-based methods. Pearson correlation was additionally computed for the multimodal correlation matrix to support comparison with prior literature in which Pearson coefficients are the convention. Statistical significance was assessed against the conventional thresholds of $p = 0.05$, 0.01, and 0.001, with three-tier asterisk notation used to indicate increasing levels of significance in summary tables.

### *B. Change-Point Detection in State Trajectory*

To complement the calendar-based phase structure used in descriptive analyses, statistically detected transition points in the biological state trajectory were identified using the Pettitt test. The Pettitt test is a non-parametric procedure designed to detect a single significant shift in the mean of a time series without requiring assumptions about the underlying distribution, making it suitable for short biological series. For each candidate split point in the series, the test statistic counts the relative ordering of values before and after the split, and the location yielding the most extreme test statistic is identified as the change-point. A p-value derived from the asymptotic distribution of the statistic is used to assess the significance of the detected transition. The test was applied independently to the Energy Entropy, Band Energy Ratio, and Motion baseline weekly series in Room 1 to identify the developmental transition weeks at which each state component exhibited a statistically significant change in mean level.

### *C. Model Stability Analysis*

The asymptotic stability of the estimated transition model was assessed through eigenvalue analysis of the state transition matrix A. Because A was constrained to diagonal form, its eigenvalues are equal to its diagonal persistence coefficients and are purely real. The spectral radius, defined as the largest absolute eigenvalue, was compared against unity: a spectral radius strictly less than one guarantees that the system is asymptotically stable, such that any deviation from the steady state decays geometrically and a unique equilibrium exists for every constant input level. This condition was verified directly from the fitted persistence coefficients.

### *D. Cross-Room Validation of the Transition Model*

To assess the generalisability of the HenTwin state transition model beyond the room in which its parameters were estimated,

a cross-room validation procedure was implemented in which the A and B parameters estimated from Room 1 data were applied to Room 5 environmental conditions and the resulting predicted state trajectory was compared against the observed Room 5 biological measurements. The validation procedure was applied to Body Surface Temperature, Energy Entropy, and Band Energy Ratio; motion was excluded from cross-room validation because video data were available for Room 1 only. Prediction accuracy was assessed using mean absolute error, root mean square error, and the coefficient of determination computed against the observed Room 5 series. A complementary direct estimation of room-specific B coefficients was conducted to determine whether observed cross-room differences reflected a difference in the magnitude of THI coupling or merely a difference in baseline level, providing diagnostic information about the structural transferability of the model.

Building on this validation, a two-tier calibration analysis was conducted to determine which model components require room-specific adjustment. Three transfer scenarios were compared, all referencing THI deviation to the Room 1 training mean to reflect a zero-local-knowledge deployment: a no-calibration scenario applying the Room 1 A and B parameters directly; a recalibration scenario retaining the Room 1 A matrix while re-estimating the input coupling and intercept on Room 5 data; and a full-refit scenario re-estimating all parameters on Room 5 as an upper bound. Transfer performance was evaluated by one-step prediction $R^2$ on Room 5 for each scenario. This one-step formulation, which predicts each state from the observed previous state, is distinct from the forward-simulation procedure used in the cross-room validation above; the two are reported separately as they characterize parameter transferability and full-trajectory generalization respectively. The persistence coefficients of the full Room 5 refit were additionally screened against the unit-stability bound to identify any non-stationary channel for which a steady state could not be meaningfully interpreted.

### *E. Principal Component Analysis*

Principal Component Analysis (PCA) was applied to the standardised multimodal feature matrix to assess the latent dimensional structure of the state vector and to verify the non-redundancy of its component branches. All variables were centred and scaled to unit variance prior to decomposition to ensure that the analysis was driven by relative variation rather than absolute scale differences across modalities. The first two principal components were retained for visualization, with score plots used to assess phase separation in the latent space and loading plots used to assess the contribution of each variable to each latent dimension.

### *F. Computational Implementation*

All data processing, statistical analyses, and state-space simulations were implemented in Python 3.9 using the pandas, NumPy, SciPy, scikit-learn, and matplotlib libraries. Optical flow estimation was performed using OpenCV 4.5 with the Dense Inverse Search algorithm. Acoustic feature extraction used the librosa library. The complete analysis pipeline was implemented in a reproducible Jupyter notebook organized into modular cells corresponding to the sections of the methodology described above. All processed datasets, fitted model parameters, perturbation simulation outputs, and figure-generation scripts are available as supplementary material, ensuring full reproducibility of the reported results.

## VII. Results

### *A. Multimodal State Vector and Cross-Modal Structure*

The finalized HenTwin state vector comprised four biological variables – body surface temperature (BST), vocalization energy entropy (EE), band energy ratio (BER), and motion baseline – with the temperature-humidity index (THI) treated as an exogenous environmental input rather than a state component. This reflects the physical causal ordering of the system, in which environmental conditions drive biological response, but biological response does not appreciably alter room-level environmental conditions over the weekly sampling interval; retaining THI within the state vector would have introduced an artificial feedback path. As shown in Fig. 6, THI declined steadily across the study in both rooms, from approximately 82 in week 1 to below 68 by week 25, with the two rooms tracking each other closely.

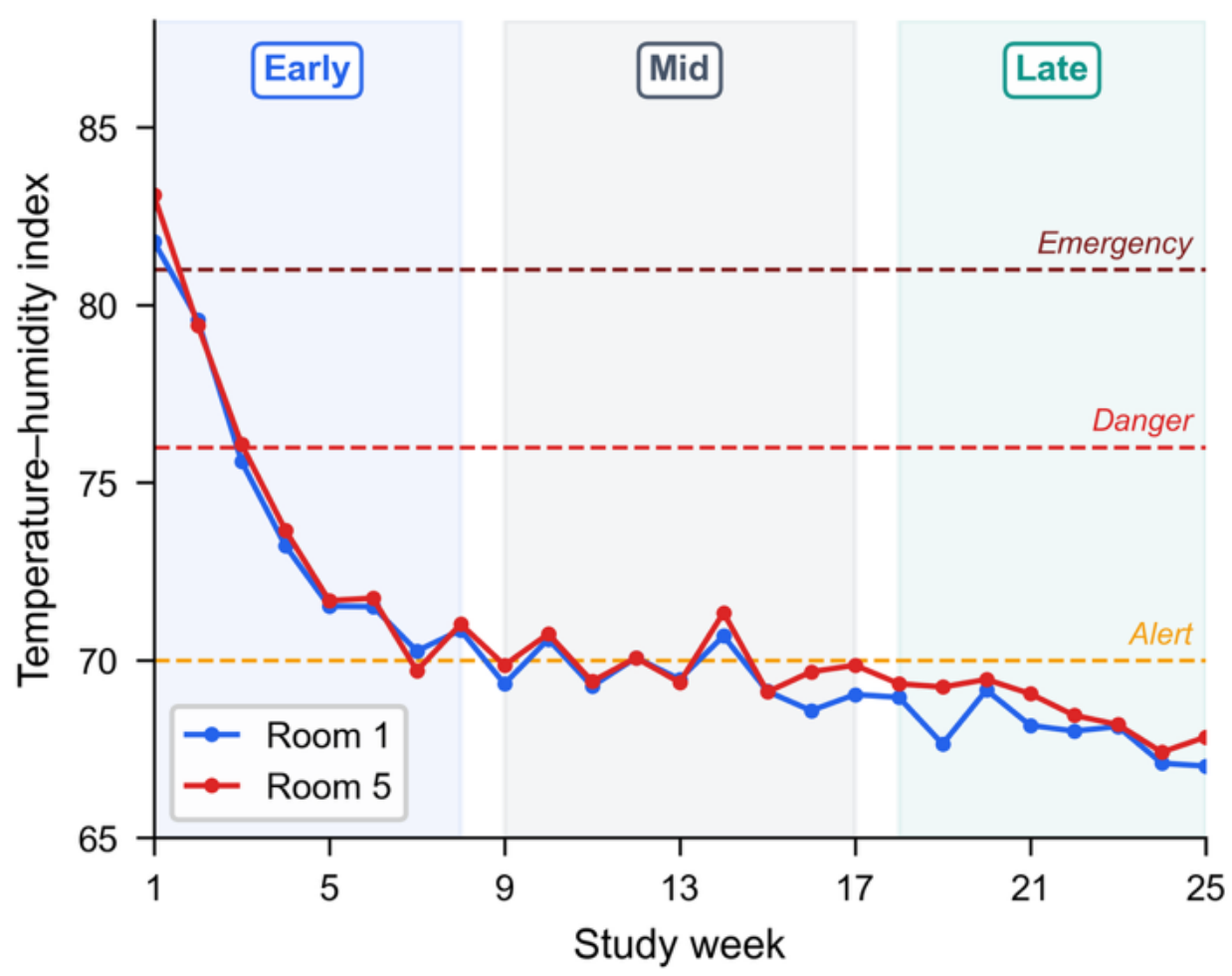

Fig. 6. Weekly temperature–humidity index (THI) for Rooms 1 and 5 across the 25-week rearing period, shown against established laying-hen thermal comfort categories (comfort < 70, alert 70–75, danger 76–81, emergency > 81) [22]. Both rooms track closely and decline steadily from approximately 82-83 in week 1 to below 68 by week 25, reflecting the graduated reduction of the brooding temperature regime. THI enters the danger and emergency range only during weeks 1-3 and from week 4 onward, both rooms remain within the comfort-to-alert range. Background shading denotes the developmental phases (early, mid, late) used in the analysis.

Spearman rank correlations in Room 1 revealed strong monotonic coupling between THI and the acoustic and thermal channels (Fig. 7). The THI–EE association was the strongest in the dataset ($\rho = 0.925$, $p < 0.001$), followed by THI–BST ($\rho = 0.806$, $p < 0.001$). THI was moderately associated with BER ($\rho = 0.565$, $p = 0.003$) and inversely associated with motion baseline (ρ = −0.594, p = 0.005), the latter indicating reduced locomotor activity under elevated thermal load. BST and EE were themselves strongly coupled (ρ = 0.835, p < 0.001), and EE was inversely related to motion baseline (ρ = −0.548, p = 0.010).

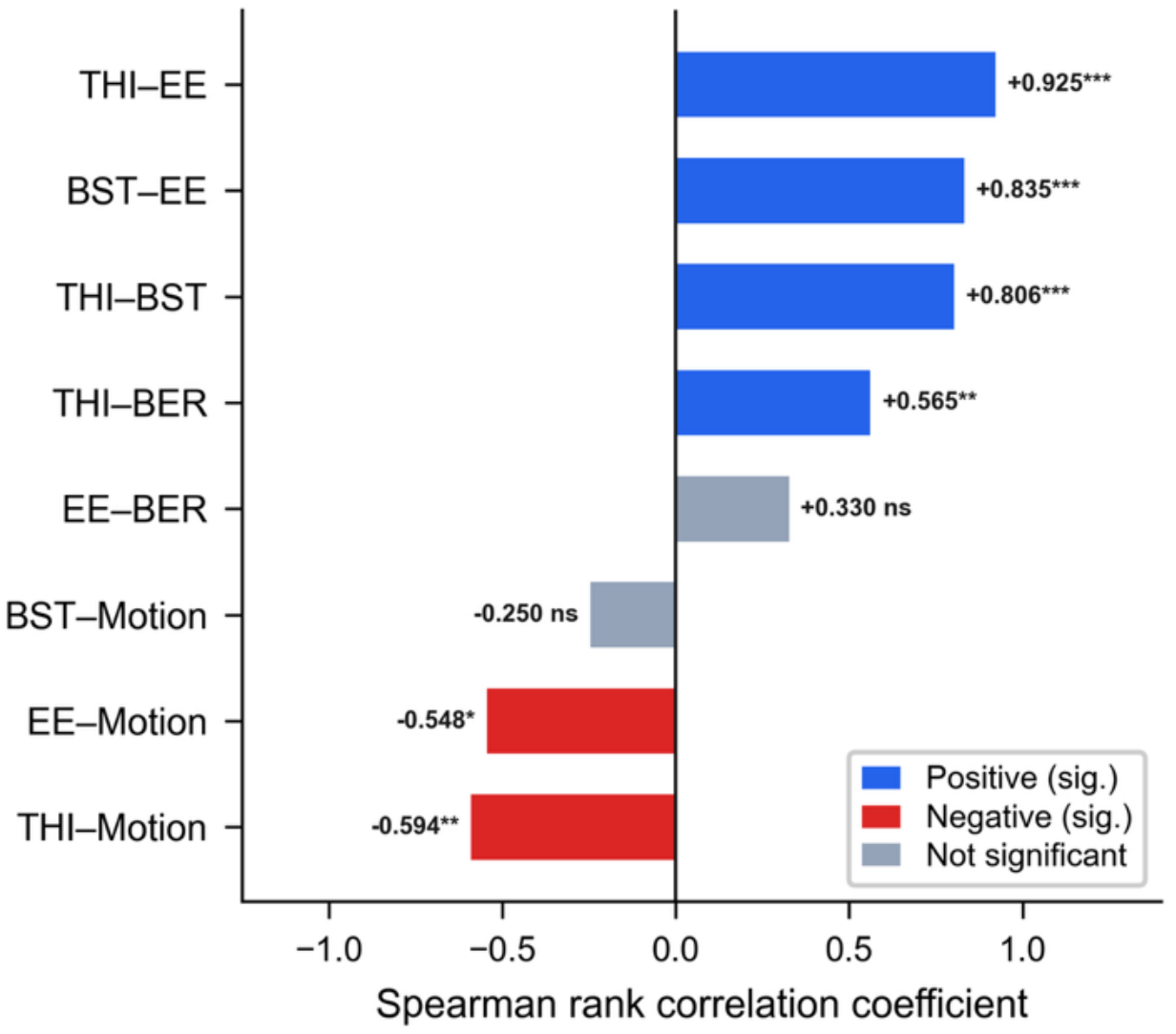

Fig. 7. Spearman rank correlations among the temperature–humidity index and the biological state variables in Room 1 (weeks 1–25). Bars are ordered by correlation coefficient and coloured by direction and significance: blue denotes significant positive coupling, red significant negative coupling, and grey non-significant associations. The temperature–humidity index is most strongly coupled to energy entropy (ρ = 0.925), followed by body surface temperature (ρ = 0.806) and band energy ratio (ρ = 0.565), and is inversely associated with motion baseline (ρ = −0.594), indicating reduced locomotor activity under elevated thermal load. The opposite-sign behaviour of the motion channel relative to the thermal and acoustic channels indicates that motion contributes non-redundant information to the state representation. Significance: *** $p < 0.001$, ** $p < 0.01$, * $p < 0.05$, ns = not significant.

As demonstrated in Fig. 8, a cross-room comparison of the principal couplings showed that the THI–BST and THI–BER associations were of comparable strength in both rooms, whereas the THI–EE coupling was substantially weaker in Room 5 (ρ = 0.484) than in Room 1 (ρ = 0.925) – a divergence examined quantitatively in Section VII.H. The consistent opposite-sign behavior of the motion channel relative to the thermal and acoustic channels indicated that motion contributed non-redundant information to the state representation.

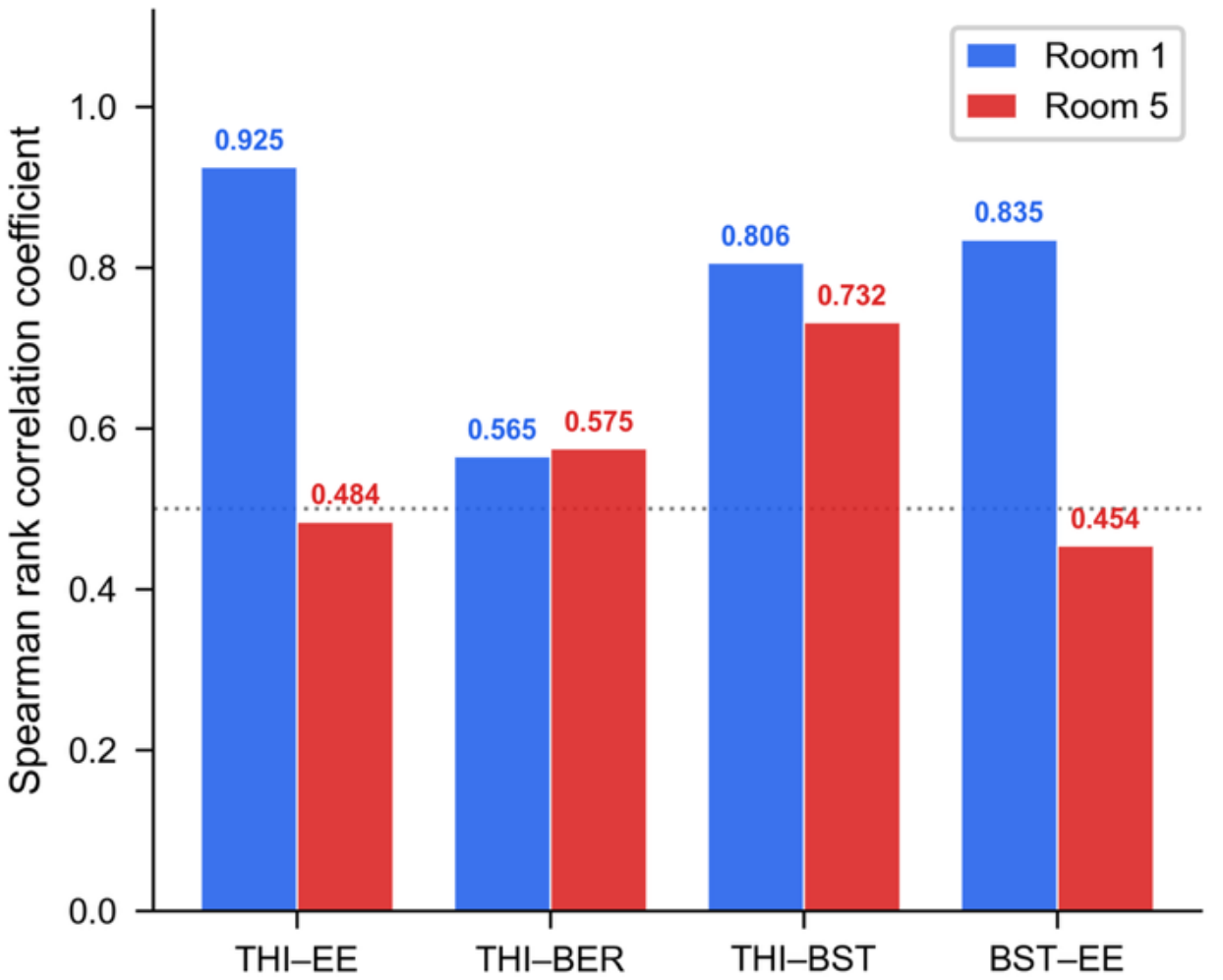

Fig. 8. Cross-room comparison of Spearman rank correlations for the principal couplings (THI–EE, THI–BER, THI–BST, and BST–EE). The THI–BST and THI–BER associations are of comparable strength in both rooms, whereas the THI–EE coupling is substantially stronger in Room 1 (ρ = 0.925) than in Room 5 (ρ = 0.484), foreshadowing the room-specific acoustic sensitivity quantified in the cross-room analysis

### *B. State Transition Model and Stability*

A discrete-time state transition model was estimated over weeks 5–25, the interval for which all four modalities were simultaneously available (20 week-pairs). The transition matrix A was constrained to diagonal form: with only 20 observations available, a full 4×4 transition matrix would have required 16 coupling parameters, guaranteeing overfitting, and preliminary attempts to fit the unconstrained matrix produced physically implausible trajectories including negative entropy values. The diagonal model, requiring four persistence parameters and four intercepts, is statistically appropriate for the available sample (Fig. 9).

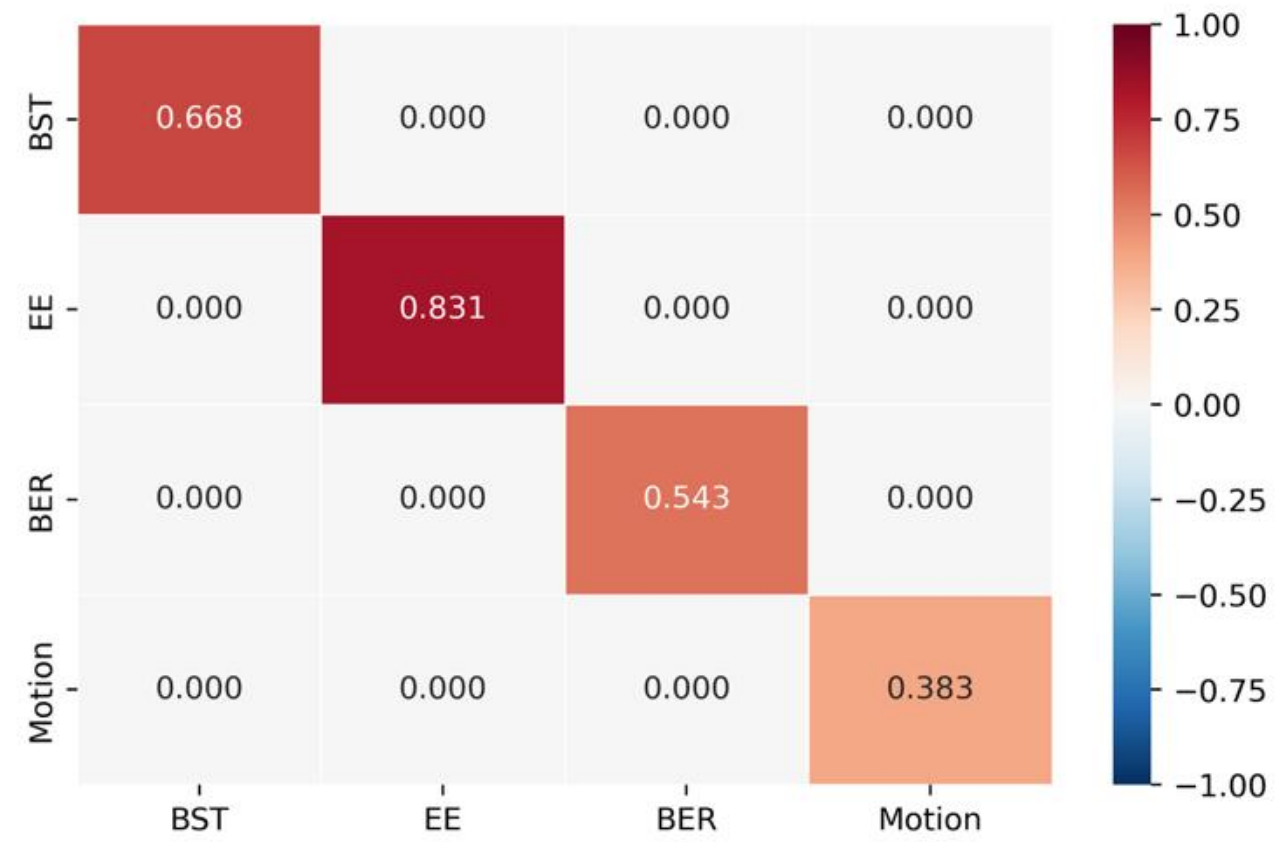

Fig. 9. Estimated state transition matrix A of the HenTwin diagonal state-space model (Room 1, weeks 5–25). The matrix is constrained to diagonal form, so each state variable evolves as a first-order autoregressive process driven by its own previous value, with cross-variable interactions absorbed into the residual term. Diagonal entries are the estimated persistence coefficients – 0.668 (BST), 0.831 (EE), 0.543 (BER), and 0.383 (motion baseline) – and all off-diagonal entries are fixed at zero by construction. Energy entropy exhibits the highest week-to-week persistence and motion baseline the lowest. All coefficients are strictly less than unity, consistent with the asymptotic stability of the model.

The estimated persistence coefficients were 0.668 (BST), 0.831 (EE), 0.543 (BER), and 0.383 (motion baseline), indicating that EE carried the strongest week-to-week memory and motion baseline the weakest. The implied steady states, computed as the fixed points c/(1−A), closely matched the corresponding observed means, confirming the internal consistency of the fitted model (Table II).

TABLE II
DIAGONAL STATE TRANSITION MODEL PARAMETERS (ROOM 1, WEEKS 5–25)

| Variable | Persistence A | Intercept c | Steady state | Observed mean | AR(1) one-step $R^2$ |
|---|---|---|---|---|---|
| BST (°C) | 0.668 | 8.280 | 24.94 | 25.03 | 0.399 |
| EE (nats) | 0.831 | 1.277 | 7.56 | 7.86 | 0.640 |
| BER (%) | 0.543 | 2.032 | 4.44 | 7.43 | 0.705 |
| Motion baseline | 0.383 | 0.359 | 0.582 | 0.572 | 0.155 |

System stability was verified directly from the diagonal structure. Because the eigenvalues of a diagonal matrix equal its diagonal entries, all four eigenvalues lie strictly inside the unit circle, with a spectral radius of 0.831 (Fig. 10). The HenTwin transition model is therefore asymptotically stable: in the absence of sustained input perturbation, any deviation from the steady state decays geometrically, and a unique equilibrium exists for every constant input level.

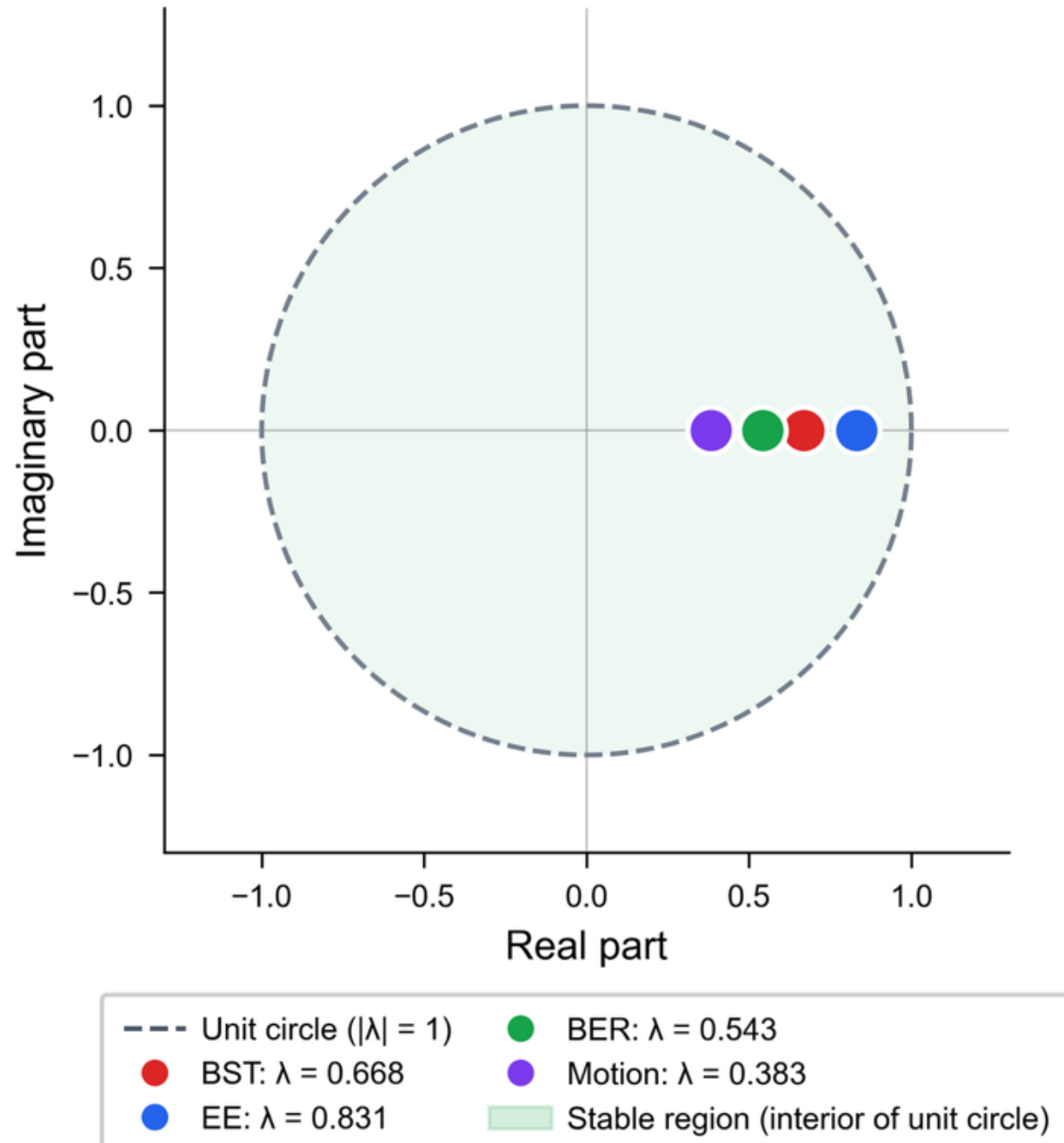

Fig. 10. Eigenvalues of the diagonal state transition matrix A plotted in the complex plane. Because A is diagonal, its eigenvalues equal its diagonal entries and are purely real: $\lambda$ = 0.668 (BST), 0.831 (EE), 0.543 (BER), and 0.383 (motion baseline). All four eigenvalues lie strictly inside the unit circle (dashed), giving a spectral radius $\rho(A) = 0.831 < 1$. The system is therefore asymptotically stable: any deviation from the steady state decays geometrically, and a unique equilibrium exists for every constant temperature–humidity input.

### C. Environmental Input Coupling

The input coupling vector B, quantifying the direct effect of a unit THI deviation on each state variable, was estimated by ordinary least squares. All four couplings were statistically significant in the direct single-variable fit. The EE coupling was the strongest, with THI deviation alone explaining 72.5% of the week-to-week variation in vocalization entropy (Table III).

TABLE III
ENVIRONMENTAL INPUT COUPLING (B VECTOR), DIRECT SINGLE-VARIABLE FIT

| Variable | B (per THI unit) | $R^2$ | p | Direction |
|---|---|---|---|---|
| BST (°C) | +0.300 | 0.462 | $< 0.001$ | rises with THI |
| EE (nats) | +0.271 | 0.725 | $< 0.001$ | rises with THI |
| BER (%) | +3.334 | 0.311 | 0.009 | rises with THI |
| Motion baseline | −0.037 | 0.327 | 0.007 | falls with THI |

The thermal-acoustic relationship is visualized in Fig. 11, which also shows the markedly weaker association in Room 5 (r = 0.52) than in Room 1 (r = 0.75), foreshadowing the room-specific sensitivity analyzed in Section VII.

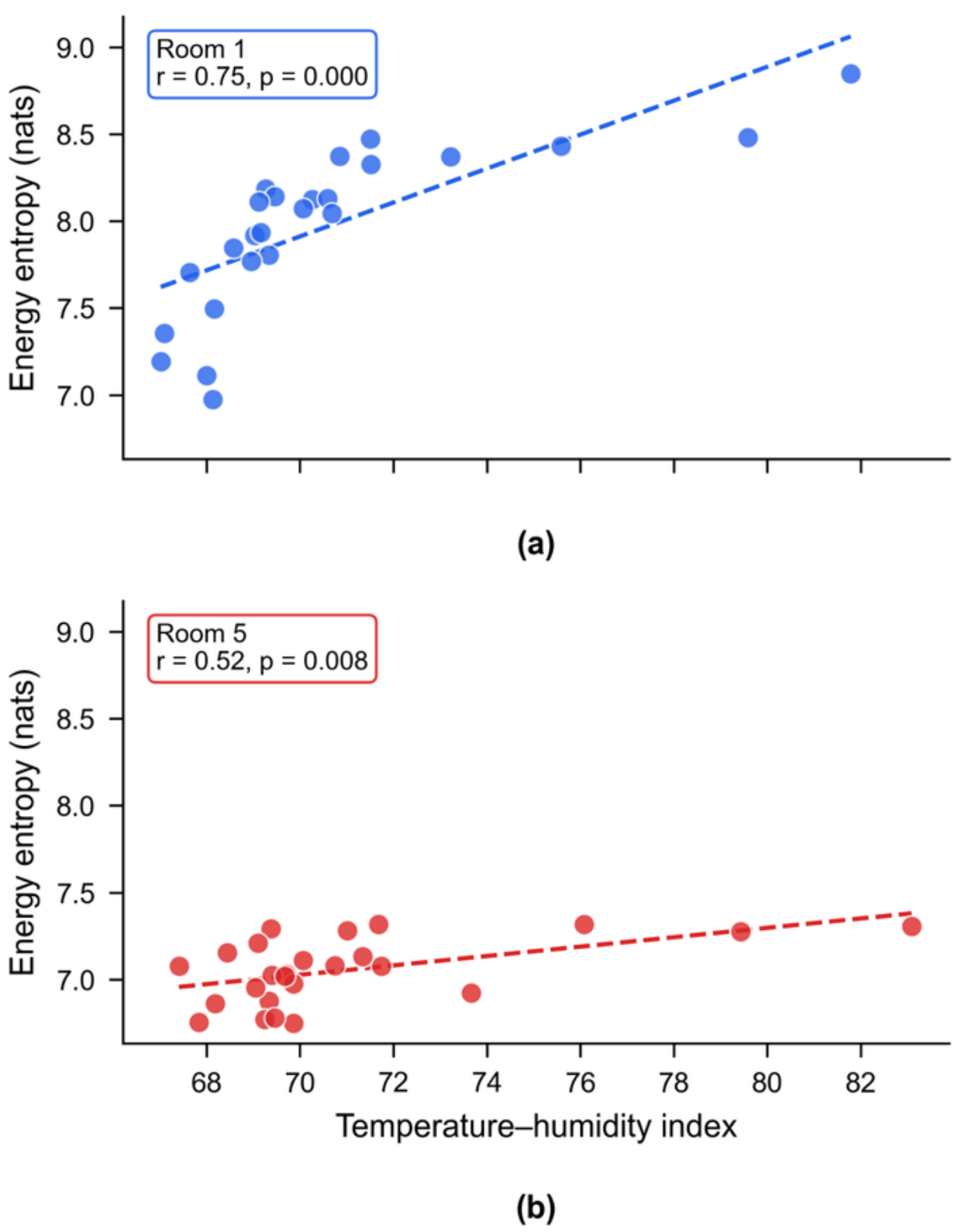

Fig. 11. Relationship between the temperature–humidity index and energy entropy across the full 25-week study (weeks 1-25), for (a) Room 1 and (b) Room 5. Each point represents one study week; dashed lines show the ordinary least squares fit, and panels share a common energy-entropy scale to allow direct comparison of slope. The association is markedly stronger in Room 1 (Pearson $r = 0.75$, $p < 0.001$) than in Room 5 ($r = 0.52$, $p = 0.008$), previewing the room-specific acoustic sensitivity to thermal load quantified in Section VII.H.

To characterize the robustness of these coefficients, the model was refitted across 2,000 bootstrap resamples. Under the BCa bootstrap, the direct single-variable couplings excluded zero for all four channels under the BCa bootstrap and were retained as the reported values. The same couplings estimated jointly with the persistence terms carried wider intervals that include zero on the 20-observation series, reflecting collinearity between each channel's previous state and the contemporaneous THI input rather than an absence of coupling. Spectral centroid, tested as a candidate fifth state variable, showed no significant THI coupling ($p = 0.23$) and, in Room 5, was almost perfectly collinear with BER ($r = 0.98$), confirming its redundancy; it was therefore excluded from the state vector. Motion reactivity, an event-locked alternative to the continuous motion baseline, was also tested and excluded as non-significant ($B = +0.017$, $R^2 = 0.096$, $p = 0.172$). The input-coupling coefficients reported here are contemporaneous structural estimates – the total association between THI deviation and each state level within the modelling window – and are used to characterize the equilibrium and perturbation behaviour of the system in Section VII.D. They are distinct from, and larger than, the marginal one-step predictive coefficients obtained when A and B are estimated jointly (Section VII.G). The contemporaneous estimate is the appropriate quantity for the steady-state and perturbation analyses but is not used for one-step forecasting.

### *D. Perturbation Analysis*

To demonstrate the intervention-aware predictive capability of the framework, a sustained perturbation of +2.0 THI units was applied from week 5 onward. This magnitude represents the upper bound of natural weekly THI variation observed in the study (range −2.26 to +2.26 units from the weekly mean) and therefore constitutes a physically realistic rather than extreme scenario. Both the baseline and perturbed simulations were initialized at the theoretical steady state of each variable in order to isolate the sustained effect of the perturbation from transient startup dynamics.

Under this perturbation, EE converged to a stable long-run increase of 0.54 nats above the unperturbed trajectory (Fig. 12). For context, the total observed developmental decline in EE across the full 25-week study was 1.87 nats. A sustained +2 THI elevation is thus predicted to shift vocalization entropy roughly a quarter of its natural developmental range. Because the modelling window spans only the thermal comfort-to-alert range (weekly mean THI 67.0-71.5, see Limitations), this perturbation probes the upper range of within-window thermal variation rather than the acute heat-stress regime of weeks 1-3, and the projected magnitudes are interpreted as structural equilibrium responses rather than as directly observed heat-stress effects.

The response across the remaining state variables showed a corresponding BST increase of approximately 0.60 °C and a motion baseline decrease of approximately 0.07 units, consistent with the signs of the estimated input coupling (Fig. 13). Band energy ratio showed a nominal increase of approximately 6.67 percentage points, but this value is not interpreted as a robust projection because the band-energy channel is non-stationary over the study period (Section VII.H).

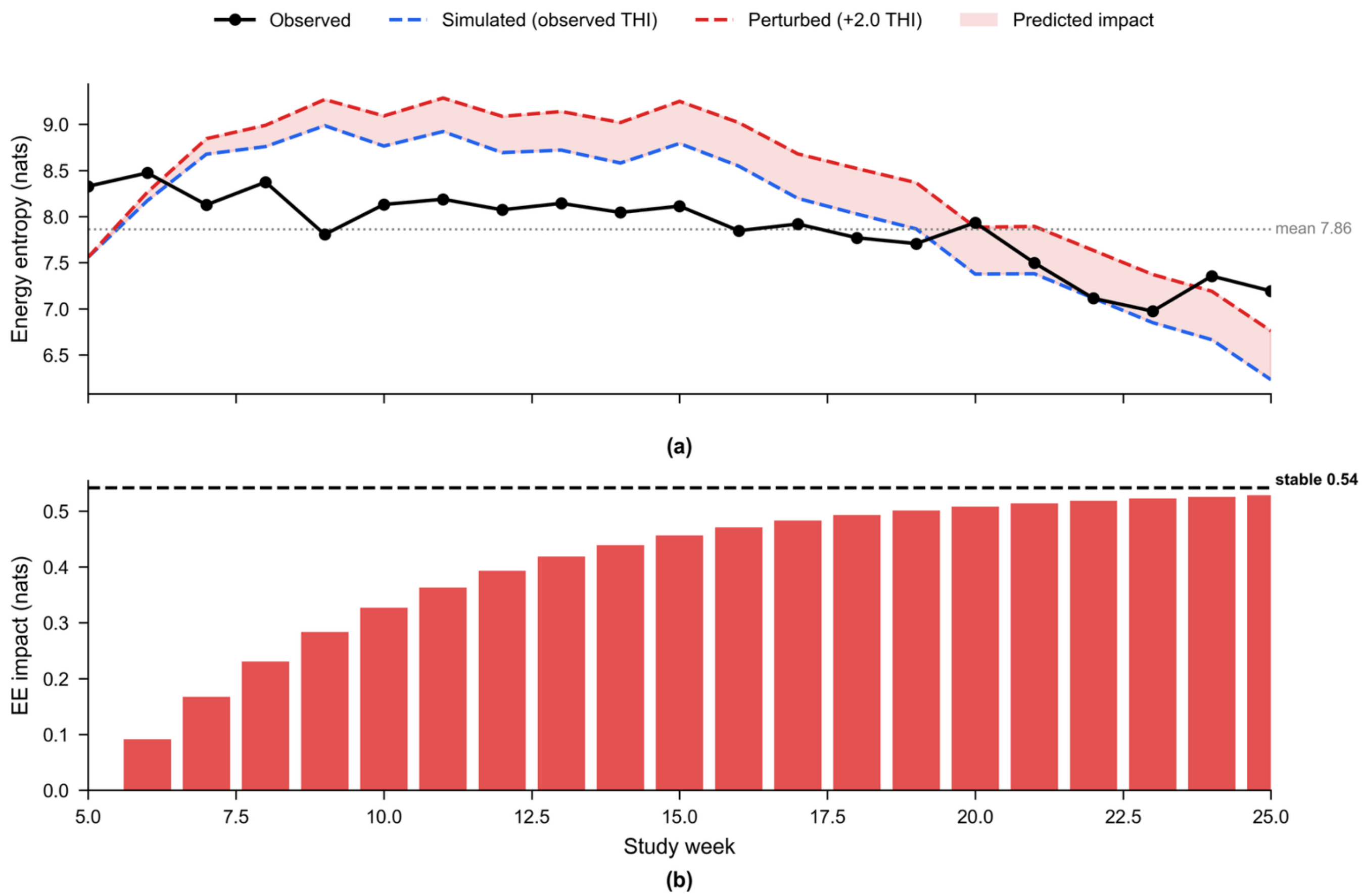

Fig. 12. Energy-entropy response to a sustained +2.0 THI perturbation, computed with the diagonal state-space model. (a) Observed weekly energy entropy (black), the model baseline trajectory under the observed THI series (blue, dashed), and the perturbed trajectory under a sustained +2.0 THI increase (red, dashed); the shaded band is the predicted perturbation impact, i.e., the difference between the perturbed and baseline trajectories. The dotted line marks the observed mean (7.86 nats). (b) Weekly perturbation impact (perturbed minus baseline), which converges to a stable long-run value of 0.54 nats – approximately one-quarter of the 1.87-nat developmental decline in energy entropy observed across the study. Both trajectories were initialized at the analytically derived steady state to isolate the sustained effect of the perturbation

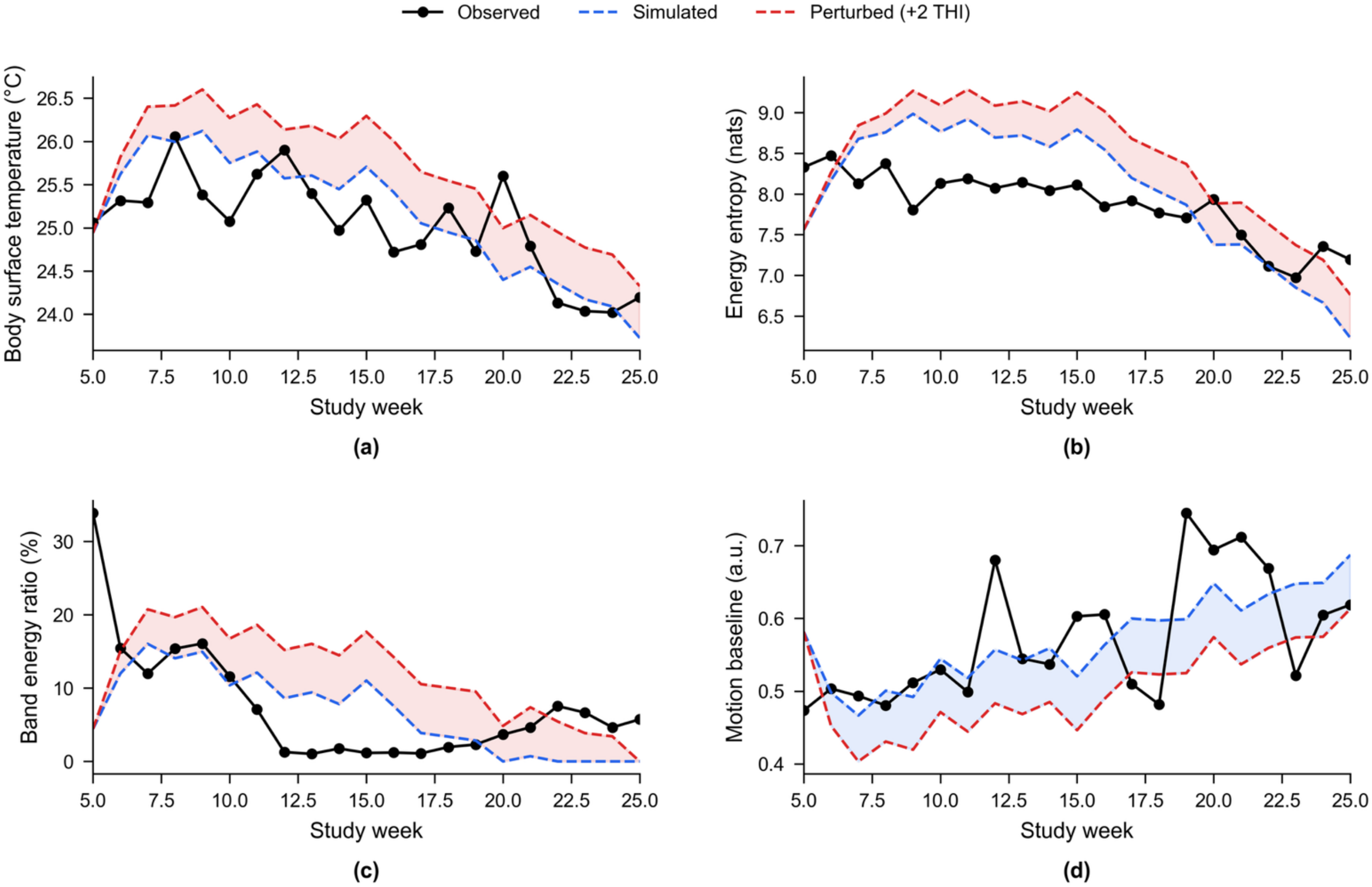

Fig. 13. Multimodal state response to a sustained +2.0 THI perturbation, computed with the diagonal state-space model. Each panel shows the observed weekly value (black), the model baseline trajectory under the observed THI series (blue, dashed), and the perturbed trajectory under a sustained +2.0 THI increase (red, dashed), with the shaded band indicating the predicted perturbation impact: (a) body surface temperature, (b) energy entropy, (c) band energy ratio, and (d) motion baseline. Body surface temperature, energy entropy, and motion baseline respond in the direction expected from the estimated input coupling — the thermal and acoustic channels increasing and motion baseline decreasing under elevated thermal load. The band-energy-ratio response (c) is shown for completeness but is not interpreted quantitatively, as that channel is non-stationary and its one-step environmental coupling is not robustly identifiable (Section VII.H). All trajectories were initialized at the analytically derived steady state to isolate the sustained effect of the perturbation.

To quantify uncertainty in these projections, the input couplings were resampled using a bias-corrected and accelerated (BCa) bootstrap with 2,000 resamples of the training weeks. All four couplings excluded zero at the 95% level, indicating individually well-determined effects despite the limited sample. The long-run EE response to a sustained +2 THI shift was estimated at +0.54 nats (95% BCa CI [+0.42, +0.69]), with corresponding intervals of [+0.32, +0.99] °C for BST and [−0.12, −0.04] units for motion baseline. These projections are interpreted as indicative of the direction and approximate magnitude of the systemic response rather than as precise forecasts, given the 20-observation training series.

### *E. Developmental Transition Detection*

The Pettitt test, a non-parametric procedure for detecting a single shift in the mean of a short time series, was applied independently to the EE, BER, and motion baseline series in Room 1 (Fig. 14). A coordinated developmental transition was detected across all three modalities within a narrow temporal window. EE exhibited a significant downward shift at week 14 ($K = 144.0$, $p < 0.001$; mean declining from 8.29 to 7.62 nats), while BER and motion baseline both shifted at week 12 (BER: $p < 0.001$; motion baseline: $K = 82.0$, $p = 0.031$, mean increasing from 0.499 to 0.609). The near-simultaneous detection of mean shifts across three independently measured modalities within weeks 12–14 indicates a system-level developmental transition rather than a modality-specific artifact.

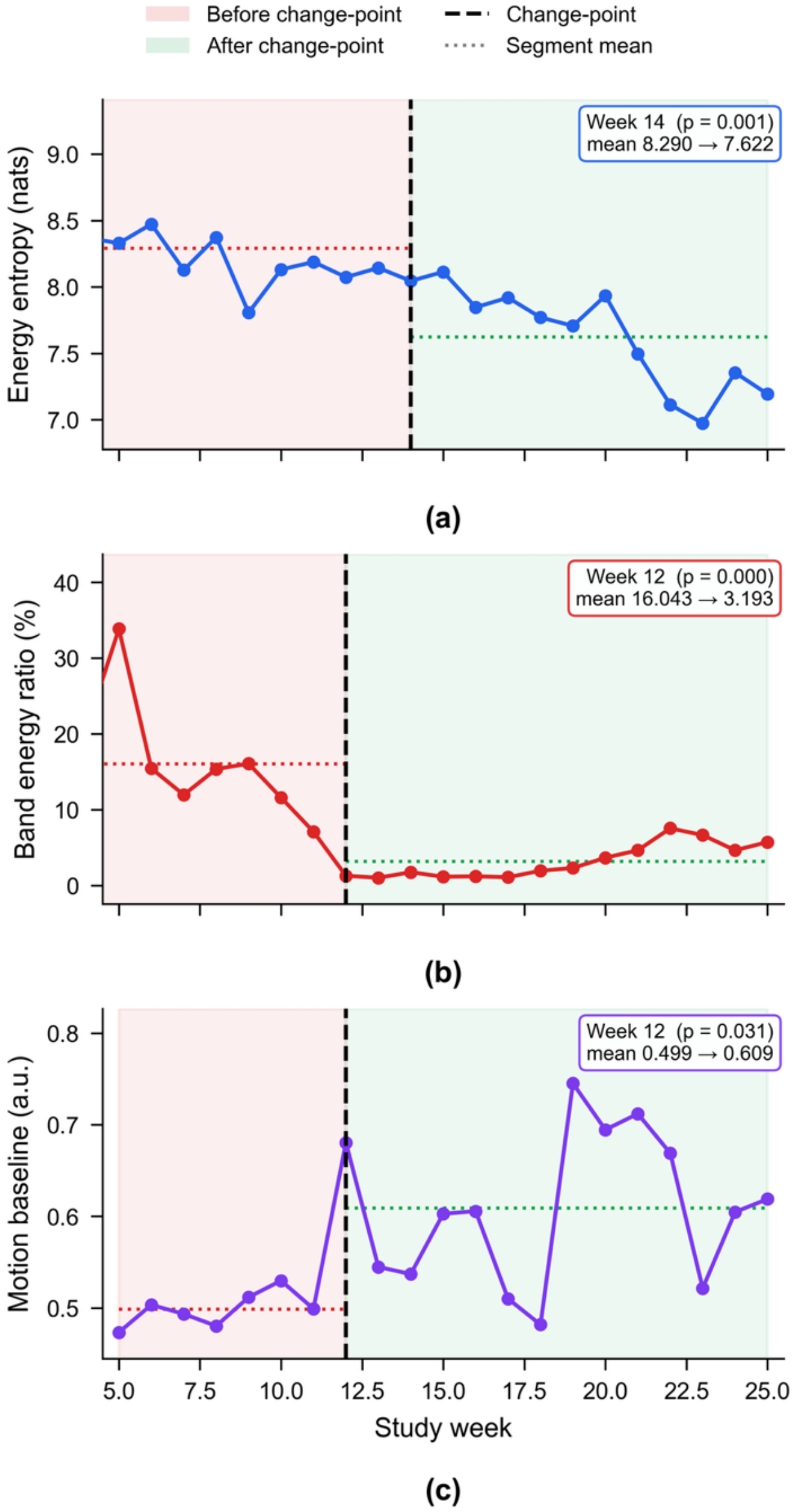

Fig. 14. Pettitt change-point detection in the Room 1 state variables: (a) energy entropy, (b) band energy ratio, and (c) motion baseline. A coordinated developmental transition is detected within a narrow window – energy entropy shifts at week 14 (K = 144, $p < 0.001$; mean 8.293 → 7.621 nats), while band energy ratio (week 12, $p < 0.001$; mean 16.043 → 3.187 %) and motion baseline (week 12, K = 82, $p = 0.031$; mean 0.499 → 0.609) shift at week 12. In each panel the dashed vertical line marks the detected change-point, dotted horizontal lines show the segment means before and after, and the shaded regions denote the pre- and post-transition periods.

### *F. Latent Dimensional Structure*

Principal component analysis of the four standardized state variables confirmed the non-redundancy of the state representation (Fig. 15). The first principal component explained 56.8% of the total variance and loaded positively on BST (0.527), EE (0.601), and BER (0.391) and negatively on motion baseline (−0.456), defining a thermal–acoustic axis along which elevated thermal and vocal activity coincide with reduced movement. The second component explained a further 27.1% of the variance and was dominated by BER (+0.628), representing a secondary behavioral dimension largely orthogonal to the primary thermal–acoustic axis. Together the two components accounted for 83.8% of the total variance. The score plot showed a separation of early, mid, and late developmental phases along PC1, consistent with the transition detected in Section VII.E, while the orthogonality of the BER and motion contributions confirmed that the behavioral channels carried information distinct from the thermal and entropy channels.

### *G. Predictive Accuracy and Cross-Validation*

One-step-ahead predictive skill was benchmarked under leave-one-out cross-validation against three baselines: a persistence (random-walk) predictor, a univariate first-order autoregressive AR(1) model, and an unconstrained vector autoregressive model with exogenous input, VARX(1), incorporating full cross-variable coupling (Table IV). For one-step forecasting the diagonal transition model was evaluated with its persistence and input-coupling parameters estimated jointly (the diagonal ARX), the configuration appropriate for short-horizon prediction; this achieved modest predictive skill, with leave-one-out coefficients of determination of 0.301,

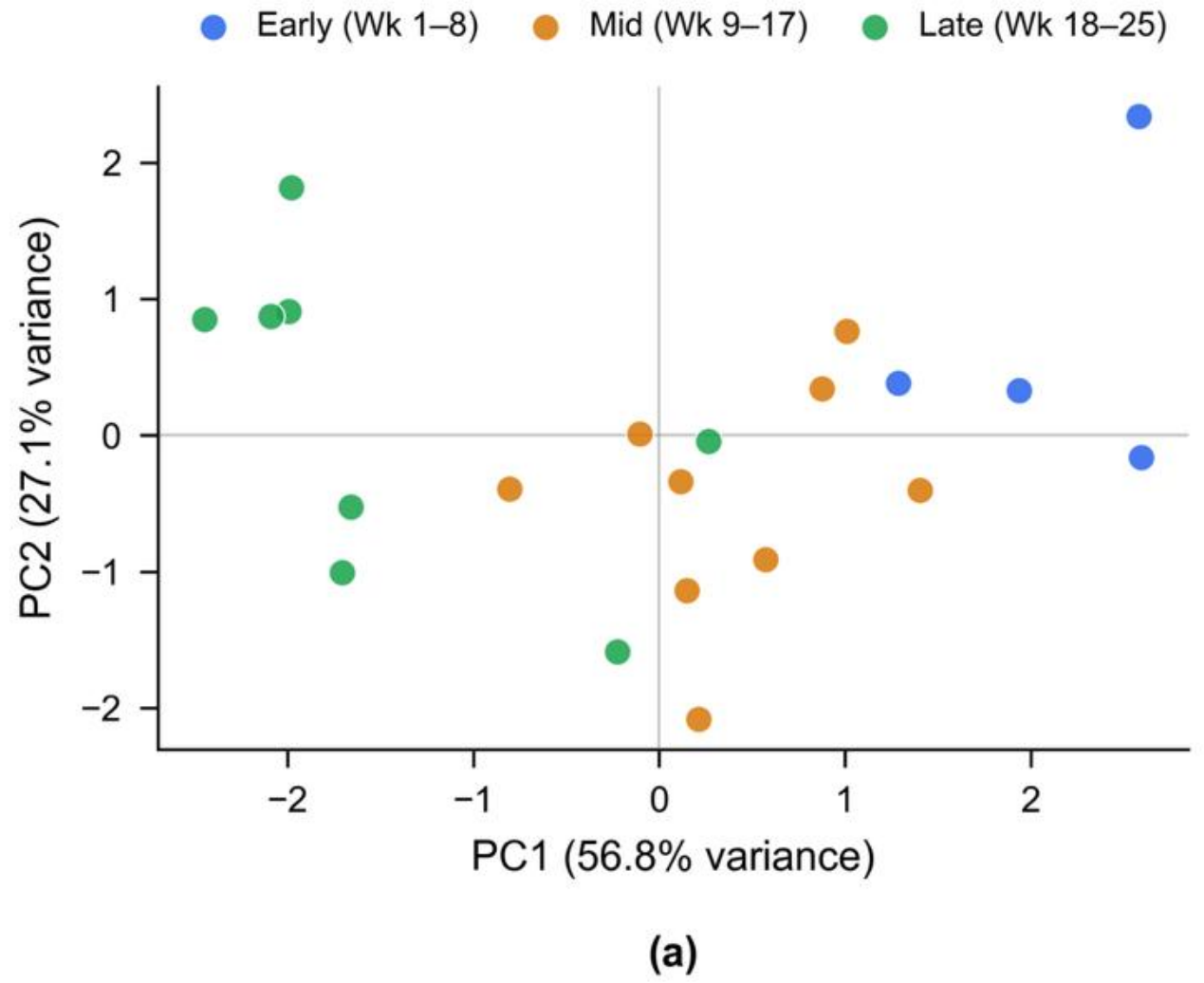

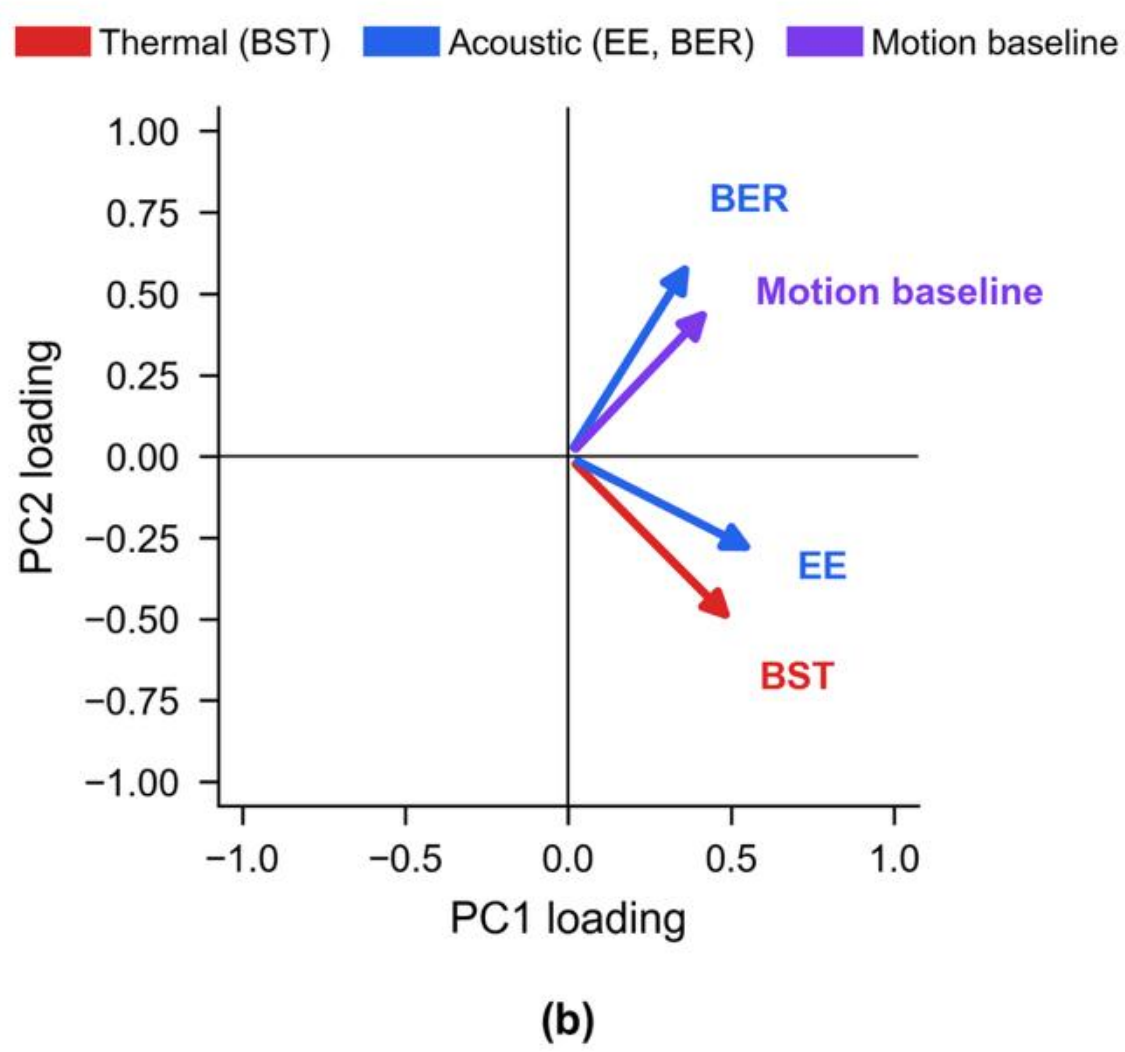

Fig. 15. Principal component analysis of the four standardised state variables (Room 1): (a) score plot and (b) loading plot. The first component (56.8% of variance) loads positively on all four variables, representing a shared developmental axis along which overall biological state changes over the rearing period. The second component (27.1%) contrasts the band-energy-ratio and motion-baseline channels (positive loadings) with body surface temperature and energy entropy (negative loadings). Together the two components account for 83.9% of the total variance, and the score plot separates the early, mid, and late developmental phases along the first component.

0.518, 0.389, and −0.003 for BST, EE, BER, and motion baseline, respectively (Table V). This skill exceeded the persistence baseline for BST and BER but not for EE, whose strong week-to-week autocorrelation rendered a random walk competitive at the one-step horizon (persistence $R^2 = 0.594$); no model resolved week-to-week motion baseline. The exogenous THI input did not materially improve one-step accuracy over the autoregressive baselines, indicating that at the weekly sampling scale the environmental signal is already largely encoded in the prior biological state. The contribution of the THI input to the framework is therefore structural – expressed through the equilibrium and perturbation behaviour characterized in Sections VII.C and VII.D – rather than through short-horizon forecasting. The structural analyses (Sections V.D–V.E, VII.C–VII.D) accordingly use the direct equilibrium couplings, while the jointly estimated couplings are used only for the one-step forecasting evaluation reported here. The unconstrained VARX(1) attained the highest in-sample fit (up to $R^2 = 0.68$) but the lowest cross-validated skill, deteriorating to $R^2 \leq 0.29$ and to −0.40 for motion baseline; this in-sample/out-of-sample divergence is the characteristic signature of overfitting and provides direct empirical support for the diagonal restriction on the transition matrix A adopted in Section VII.B.

TABLE IV
ONE-STEP-AHEAD FORECAST SKILL VERSUS BASELINES (ROOM 1, WEEKS 5-25, LEAVE-ONE-OUT $R^2$)

| Variable | Persistence (RW) | AR(1) | Diagonal ARX | Full VARX |
|---|---|---|---|---|
| BST (°C) | 0.295 | 0.285 | 0.301 | 0.233 |
| EE (nats) | 0.594 | 0.536 | 0.518 | 0.294 |
| BER (%) | 0.129 | 0.413 | 0.389 | 0.240 |
| Motion baseline | -0.255 | 0.055 | -0.003 | -0.403 |

TABLE V
ONE-STEP-AHEAD PREDICTION ACCURACY OF THE DIAGONAL ARX MODEL (ROOM 1, WEEKS 5-25, LEAVE-ONE-OUT $R^2$)

| Variable | MAE | RMSE | $R^2$ (LOO) |
|---|---|---|---|
| BST (°C) | 0.393 | 0.489 | 0.301 |
| EE (nats) | 0.239 | 0.283 | 0.518 |
| BER (%) | 2.644 | 4.000 | 0.389 |
| Motion baseline | 0.065 | 0.083 | −0.003 |

One-step residual diagnostics were recomputed for the jointly estimated diagonal ARX model (Fig. S1). Residuals for energy entropy, band energy ratio, and motion baseline were approximately normally distributed (Shapiro–Wilk p = 0.44, 0.07, and 0.37), while body surface temperature residuals showed a marginal departure from normality ($p = 0.041$), driven largely by the early developmental weeks. Autocorrelation diagnostics indicated that body surface temperature and motion baseline residuals were consistent with white noise, with no significant autocorrelation under the Ljung–Box test ($p = 0.78$ and 0.61 at lag 4) and Durbin–Watson statistics of 2.30 and 2.02. Energy-entropy residuals likewise showed no significant Ljung–Box autocorrelation ($p = 0.43$); their Durbin–Watson statistic of 2.61 fell within the inconclusive region for $n = 20$ and two regressors (bounds ≈ 1.10 and 1.54), indicating at most mild negative lag-one autocorrelation. Band-energy-ratio residuals exhibited significant positive autocorrelation (Durbin–Watson = 0.94) and a borderline Ljung–Box result ($p = 0.053$), consistent with the trending, non-stationary behaviour of that channel noted in Section VII.H. Because the training series comprises only 20 observations, both the Ljung–Box and Shapiro–Wilk tests have limited power; these results therefore indicate the absence of gross residual structure rather than strong confirmation of ideal residual behaviour.

Impulse response analysis of the fitted diagonal transition model (Fig. S2) characterized the intrinsic temporal persistence of each channel – that is, the decay of a unit THI impulse under the autoregressive persistence coefficients of Section VII.B, independent of the forecasting parameterization above. The channel-specific half-lives were approximately 3.7 weeks for energy entropy, 1.7 weeks for body surface temperature, 1.1 weeks for band energy ratio, and 0.7 weeks for motion baseline. Energy entropy thus retains environmental effects longest, consistent with its high persistence coefficient and its role as the slowest-adapting channel in the state vector.

### *H. Cross-Room Transferability and Two-Tier Calibration*

The generalizability of the Room 1 transition model was assessed by applying its estimated parameters to Room 5 environmental conditions and comparing the predicted trajectory against observed Room 5 measurements (Fig. 16). Motion was excluded from this analysis because video data were available for Room 1 only.

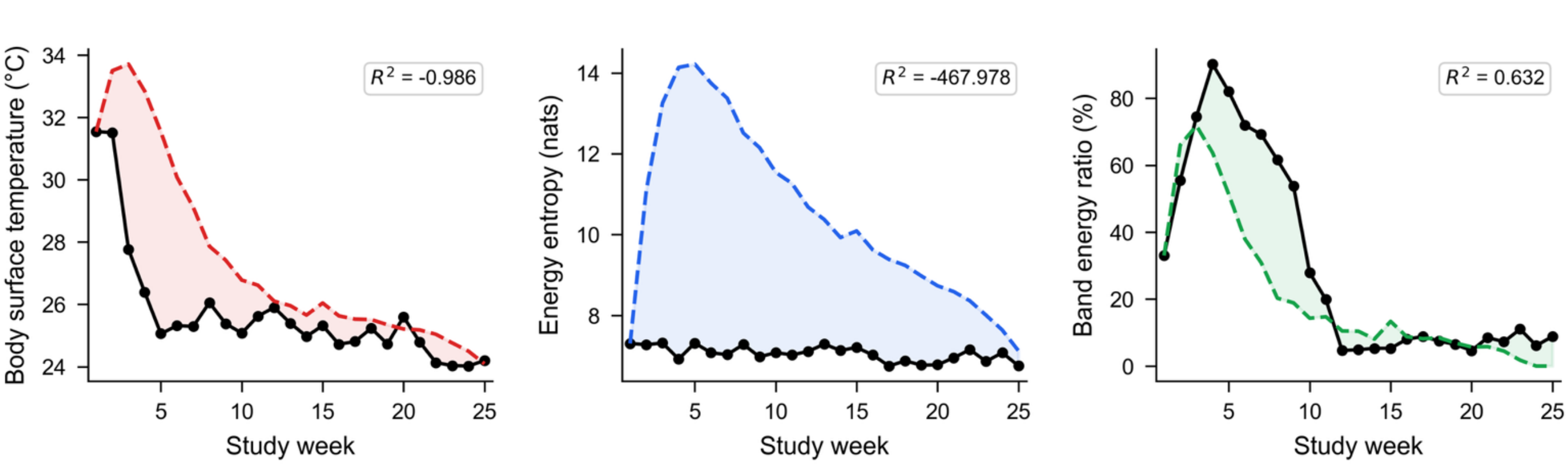

Fig. 16. Direct cross-room transfer of the Room 1 state-space model applied, without recalibration, to Room 5 observations: (a) body surface temperature, (b) energy entropy, and (c) band energy ratio. In each panel the Room 5 observed series (black) is compared with the Room 1 model prediction (dashed), with the shaded band indicating the prediction error and the leave-one-out coefficient of determination annotated. Transfer is poor for body surface temperature ($R^2 = -0.986$) and fails for energy entropy ($R^2 = -467.978$), where the Room 1 model greatly overpredicts Room 5 entropy, whereas band energy ratio transfers moderately well ($R^2 = 0.632$). The collapse of the energy-entropy prediction reflects the markedly weaker thermal–acoustic coupling in Room 5 and motivates the room-specific recalibration described in Section VII.H.

Transferability was strongly variable-dependent (Table VI). BER generalized well across rooms ($R^2 = 0.632$), indicating that the band-energy structure of vocalizations responds to thermal load in a conserved manner. BST could not be meaningfully validated ($R^2 = -0.986$), a direct consequence of the farm-wide thermal imaging limitation, whereby a single non-room-tagged BST series was applied to both rooms. EE failed to transfer under direct application of the Room 1 model ($R^2 = -467.98$).

TABLE VI
CROSS-ROOM VALIDATION: ROOM 1 MODEL APPLIED TO ROOM 5

| Variable | MAE | RMSE | $R^2$ | Observed mean | Predicted mean |
|---|---|---|---|---|---|
| BST (°C) | 1.748 | 2.657 | −0.986 | 25.71 | 27.42 |
| EE (nats) | 3.401 | 3.994 | −467.98 | 7.05 | 10.45 |
| BER (%) | 11.740 | 17.753 | +0.632 | 29.47 | 20.61 |

Direct estimation of room-specific EE coupling clarified the source of this failure (Fig. 17). The THI–EE coupling in Room 5 (B = 0.027 nats per THI unit, $R^2 = 0.271$, $p = 0.008$) was an order of magnitude weaker than in Room 1 (B = 0.271, $R^2 = 0.725$, $p < 0.001$), and the two rooms additionally differed by a persistent baseline offset of 0.917 nats, with Room 1 exhibiting consistently higher entropy throughout the study. The cross-room EE failure therefore reflects both a difference in coupling magnitude and a difference in baseline level rather than a defect in the model structure.

A two-tier calibration analysis quantified the remedy (Fig. 18, Table VII). When the Room 1 transition structure (A matrix) was retained and only the input coupling and intercept were recalibrated on Room 5 data, one-step EE prediction recovered from $R^2 = -39.50$ under direct transfer to $R^2 = -0.14$, a gain of 39.4 $R^2$ units; recalibration similarly improved BST (gain +0.54) and left the already-strong BER transfer essentially unchanged (gain +0.02). The recalibrated EE coupling was near zero (0.004 nats per THI unit), indicating that the dominant cross-room discrepancy in the entropy channel is attributable to the persistent baseline offset rather than to differential THI sensitivity, and that intercept correction is consequently the critical calibration step.

TABLE VII
TWO-TIER TRANSFER TEST: ONE-STEP $R^2$ ON ROOM 5

| Variable | No calibration | Recal. B + intercept | Full Room 5 refit | Gain (recal − none) |
|---|---|---|---|---|
| BST | 0.238 | 0.782 | 0.627 | +0.544 |
| EE | −39.50 | −0.139 | 0.215 | +39.36 |
| BER | 0.874 | 0.899 | 0.674 | +0.024 |

The Room 5 BER persistence coefficient exceeded unity under full refitting (A = 1.13), indicating non-stationarity of the BER series over the full study period; BER transferability is therefore reported through one-step $R^2$ only, and its steady state is not interpreted. Taken together, these results indicate that the transition structure is portable across rooms while the input sensitivity and baseline require room-specific calibration – supporting a two-tier deployment architecture comprising a shared structural model and a lightweight per-room calibration stage.

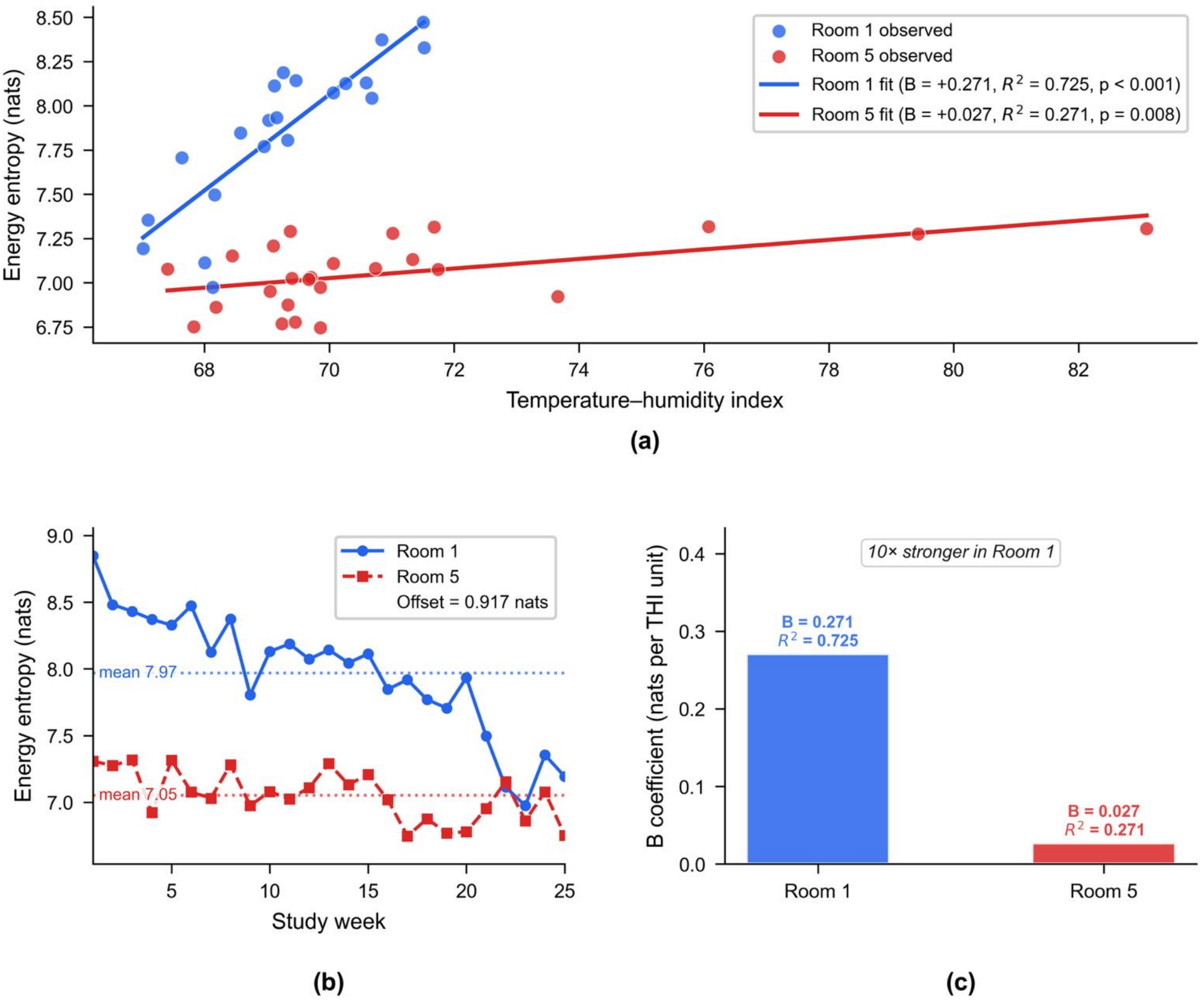

Fig. 17. Between-room acoustic sensitivity to thermal load. (a) Energy entropy versus the temperature–humidity index for Room 1 (blue) and Room 5 (red), with ordinary least squares fits; the THI–EE coupling is approximately an order of magnitude stronger in Room 1 (B = +0.271 nats per THI unit, $R^2 = 0.725$, $p < 0.001$) than in Room 5 (B = +0.027 nats per THI unit, $R^2 = 0.271$, $p = 0.008$). (b) Weekly energy-entropy trajectories for the two rooms across the full 25-week study, showing a persistent baseline offset of 0.917 nats, with Room 1 maintaining consistently higher entropy throughout (room means 7.97 and 7.05 nats, dotted lines). (c) Direct comparison of the THI–EE coupling coefficient between rooms, illustrating the order-of-magnitude difference. Together the panels show that the cross-room divergence in the acoustic channel reflects both a weaker THI coupling and a persistent baseline offset in Room 5, rather than a failure of the model structure.

## VIII. Discussion

### A. Principal Contributions

This work advances the HenTwin framework from a multimodal monitoring pipeline toward a formal state-space representation of the multimodal biological state of a laying-hen flock. Five elements characterize the contribution. First, the biological state is represented by an explicit four-dimensional state vector with environmental conditions modeled as an exogenous input, establishing a clear causal architecture and a verifiably stable transition operator. Second, a discrete-time transition model with estimated persistence and environmental coupling parameters provides a generative description of how the system evolves, rather than a static correlational summary. Third, a perturbation simulation projects the systemic consequences of a sustained environmental change, demonstrating intervention-aware prediction. Fourth, a cross-room validation and two-tier calibration analysis characterize the structural transferability of the model across deployment sites. Fifth, a leave-one-out cross-validation provides a generalization-aware assessment of predictive accuracy.

### B. Interpretation of the System Dynamics

The estimated dynamics are physically coherent. The high persistence of the entropy channel and its strong contemporaneous coupling to thermal load make EE the most environmentally responsive channel in the state vector. A sustained thermal shift produces a measurable long-run entropy elevation, modest in magnitude relative to the developmental range but consistent in direction across the multimodal state. The impulse response analysis makes this explicit, with EE showing the longest half-life of any channel (3.7 weeks). The inverse coupling of the motion channel, in which locomotor activity declines as thermal and acoustic activity rise, is consistent with the reduction of voluntary movement under thermal load and supports the inclusion of motion as a non-redundant dimension of the state vector, a conclusion reinforced by the orthogonal loading of the behavioral channels in the principal component analysis.

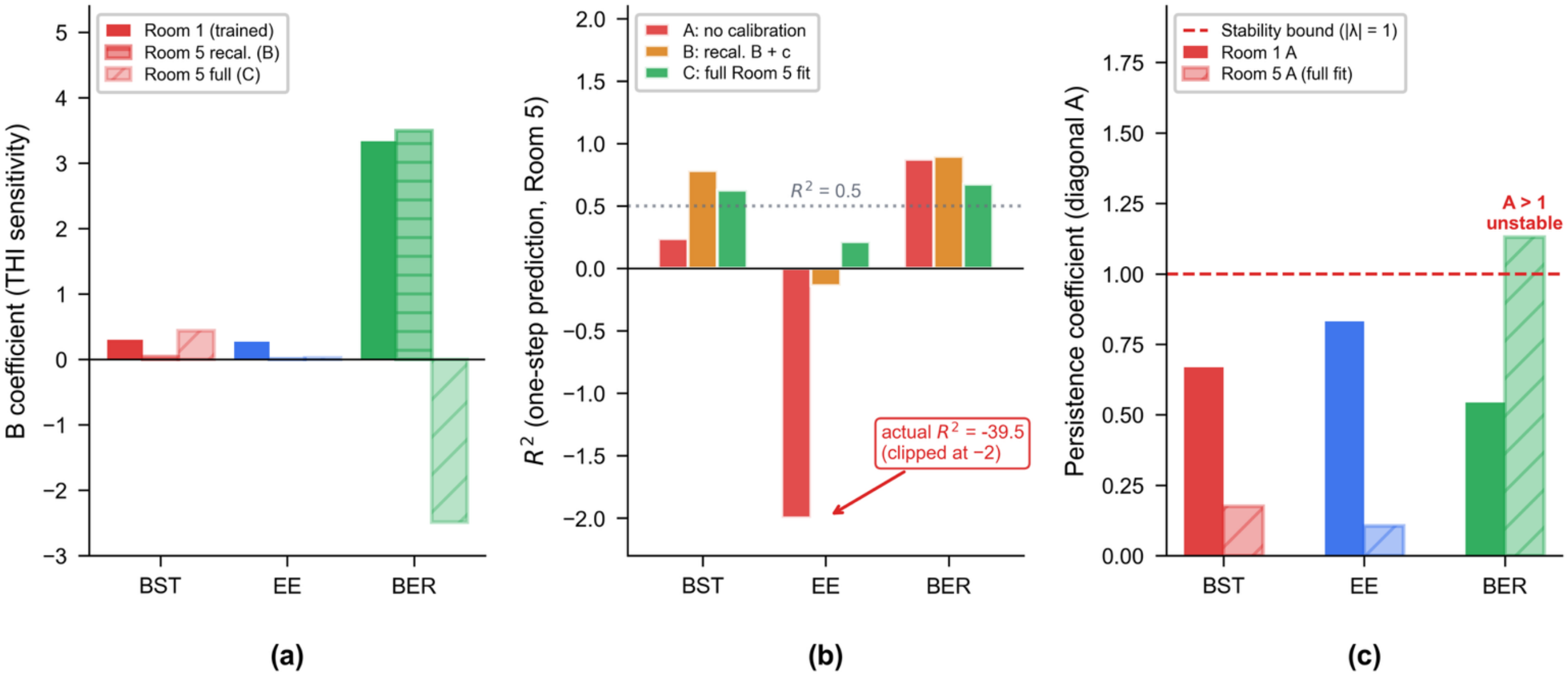

Fig. 18. Two-tier calibration transfer test, Room 1 model applied to Room 5 (all 25 weeks; THI deviation referenced to the Room 1 training mean). (a) Input coupling coefficients (B) for each shared state variable under three conditions – the trained Room 1 vector, the Room 5 recalibrated vector (Scenario B, intercept and B refitted while the Room 1 A matrix is retained), and the full Room 5 refit (Scenario C). The recalibrated EE coupling collapses to near zero, indicating that the cross-room discrepancy in the entropy channel is governed by the persistent baseline offset rather than by differential THI sensitivity. (b) One-step prediction $R^2$ on Room 5 under the three scenarios; the axis is clipped at −2 for legibility and the actual no-calibration EE value of −39.5 is annotated. Recalibrating the intercept and B (Scenario B) recovers EE from $R^2 = -39.5$ to −0.14 and improves BST, while the already-strong BER transfer is essentially unchanged; the dotted line marks $R^2 = 0.5$. (c) Persistence coefficients (diagonal of A) for Room 1 and the full Room 5 refit, with the stability boundary $|\lambda| = 1$ (dashed). BST and EE remain comfortably within the stable region in both rooms, though Room 5 shows lower persistence than Room 1. The Room 5 BER coefficient exceeds unity (A ≈ 1.13, flagged), indicating non-stationarity of the BER series over the full study period; BER transfer is therefore reported through one-step $R^2$ only and its steady state is not interpreted.

The coordinated mean shift detected across three modalities within weeks 12–14 is interpreted as a developmental transition of the system state rather than a welfare event. The simultaneity of the shifts across independently measured channels is the key observation: it indicates that the transition is a property of the underlying system rather than of any single sensing modality, and it coincides with the phase separation visible along PC1 in the principal component analysis.

### C. Transferability and Deployment Implications

The cross-room analysis yields a result of direct relevance to scalable IoT deployment. The finding that the transition structure transfers across rooms while the input coupling and baseline does not imply that a HenTwin instance deployed to a new room does not require full re-estimation. Rather, a short local observation period sufficient to recalibrate the input coupling and intercept is adequate. The magnitude of this recovery was substantial: retaining the Room 1 transition structure and recalibrating only the input coupling and intercept improved one-step energy entropy prediction on Room 5 from $R^2 = -39.50$ to −0.14, with the recalibrated coupling falling to near zero and confirming that intercept correction, rather than re-estimation of thermal sensitivity, is the operative calibration step. This calibration stage must additionally screen each channel for non-stationarity, as the Room 5 band energy ratio exhibited a persistence coefficient exceeding unity under full refitting and therefore admitted no stable steady state for transfer. The order-of-magnitude difference in acoustic THI sensitivity between the two rooms is itself a substantive empirical finding, suggesting that the acoustic response of a flock to thermal load is strongly conditioned by room-specific factors – plausibly including microphone placement, room geometry, and local acoustic properties – and that acoustic channels in particular should not be assumed transferable without local calibration. The two-tier architecture that follows from this result – a shared structural model coupled with a lightweight per-room calibration stage – is a practical template for deploying the framework across a multi-room facility.

### D. Limitations

Several limitations bound the interpretation of these results. The training series comprised 20 week-pairs from a single rearing cohort, which both necessitated the diagonal restriction on the transition matrix and produced the wide bootstrap intervals on the perturbation projections; the projections should accordingly be read as indications of direction and approximate magnitude rather than as precise forecasts. The weekly temporal resolution limits the model's week-to-week predictive accuracy, as reflected in the modest leave-one-out coefficients of determination. The thermal imaging data were acquired at the farm level rather than per room, which prevented meaningful cross-room validation of the BST channel and represents the principal sensing limitation of the study. The present model also departs from a complete state-space formulation in that it omits an explicit observation equation and operates in batch rather than in real time; it is therefore a discrete-time transition model with exogenous input rather than a full recursive estimator.

A scoping limitation arises from the modality-availability window. Because behavioural (optical-flow) features were available only from week 5, the transition model, perturbation analysis, change-point detection, and dimensional analysis were estimated over weeks 5-25. Reconstruction of the weekly temperature–humidity index shows that this window falls within the thermal comfort-to-alert range for laying hens (weekly mean THI 67.0-71.5), whereas the acute heat-stress episode of the study weeks 1-3, with weekly mean THI of 81.8, 79.6, and 75.6, spanning the danger and emergency categories, precedes the multimodal window and is therefore characterized structurally, through the fitted input coupling and perturbation response, rather than observed directly in the state vector. The +2.0 THI perturbation of Section VII.D accordingly probes the upper range of within-window thermal variation rather than the acute heat-stress regime, and its projected magnitudes should be read as structural equilibrium responses extrapolated modestly beyond the observed comfort-band envelope. Direct multimodal characterization of the acute heat-stress phase would require behavioural instrumentation from week 1 and is identified as a priority for subsequent deployments.

The framework presently models multimodal biological state and its response to environmental conditions. Terms such as developmental transition and system state are therefore interpreted within the context of the inferred biological state representation rather than as direct welfare assessments. Establishing explicit relationships between the inferred multimodal state representation and independently validated welfare outcomes remains an important direction for future research.

### E. Future Work

Three developments would directly address the limitations above. Increasing the temporal resolution from weekly to daily sampling would expand the training series to approximately 75 observations, rendering regularized vector autoregressive structures with off-diagonal coupling statistically feasible. Acquiring a second rearing cohort would provide an independent dataset for external validation and would allow the persistence and coupling parameters to be tested for stability across cohorts. Finally, sensor calibration enabling an explicit observation model would permit the formulation of a full recursive state estimator with real-time updating, completing the transition from the present batch transition model toward an operational digital twin. Integrating independent welfare indicators alongside these methodological developments would allow the multimodal state representation to be validated against established welfare outcomes.

## IX. Conclusion

HenTwin advances precision-livestock monitoring from multimodal measurement toward formal state-space inference of a flock's multimodal biological state. By defining a system state vector that integrates thermal, acoustic, and activity components with environmental conditions as an exogenous input, formulating a verifiably stable discrete-time transition model that captures both developmental dynamics and environmental forcing, and demonstrating intervention-aware perturbation analysis, this work establishes a scalable foundation for state-aware decision support in precision livestock systems. The asymptotic stability of the estimated transition operator, the coordinated developmental transition detected across three independent modalities, the portability of the transition structure across rooms under lightweight per-room calibration, and the demonstrated response to sustained THI perturbation collectively distinguish HenTwin from descriptive monitoring pipelines and move it toward an operational digital twin. Realizing that goal fully will require sub-week temporal resolution, multi-cohort validation, and an explicit observation model enabling real-time recursive state estimation; integrating independent welfare indicators alongside these developments would allow the multimodal state representation to be validated against established welfare outcomes.

## Acknowledgment

The authors thank the staff of the Atlantic Poultry Research Centre for their support throughout this study. The authors are also grateful to Daniel Essien, PhD student (Computer Science), Dalhousie University, for his assistance with video analysis and feature extraction, and to Jean Lynds, Farm Manager at the Dalhousie Agricultural Campus, and Michael McConkey, Programs and Finance Manager-Farm, Dalhousie Agricultural Campus, for their assistance and cooperation.

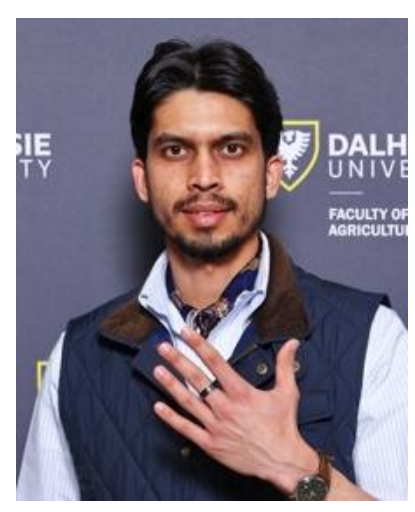

**Yashan Dhaliwal** (Undergraduate Student Member, IEEE) received the B.Sc. degree in bioveterinary science (honours) with a minor in equine science from Dalhousie University, Agricultural Campus, Truro, NS, Canada, in 2026.

He is currently a Research Assistant with the MooAnalytica Lab, Dalhousie Agricultural Campus, Truro, NS, Canada, where his work focuses on animal welfare and precision livestock farming (PLF). His undergraduate research experience includes AI-based early detection of lameness in dairy cows, multimodal welfare monitoring of laying hens, and the development of big-data infrastructure for precision poultry systems. He is currently contributing to HenTwin, a digital twin project for poultry, while also beginning research on assessing equine stress, latent states, and emotions using multimodal sensing technologies. His research interests include precision livestock farming, multimodal sensing for animal welfare, and the integration of veterinary science with emerging sensing technologies.

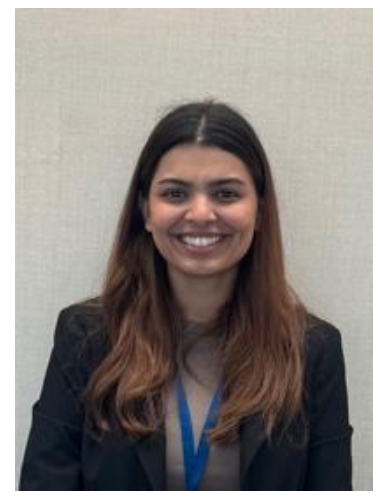

**SHREYA RAO** received the B.Tech. degree in computer science from MKSSS's Cummins College of Engineering for Women, Pune India in 2021, and is currently pursuing the M.C.S. degree in computer science at Dalhousie University, Halifax, NS, Canada.

She is a member of the Mooanalytica research laboratory under the supervision of Prof. Suresh Neethirajan. Her research focuses on the integration of computer vision, state-space modelling, and cyber-physical systems for precision livestock farming, with particular emphasis on digital twin frameworks for dairy cattle and poultry welfare monitoring. She has contributed to the development of multimodal sensing pipelines, behaviour-driven state-space digital twin architectures, and IoT-based welfare inference systems. Her work is conducted at the Ruminant Animal Centre and the Atlantic Poultry Research Centre at the Dalhousie Agricultural Campus, Truro, NS, Canada.

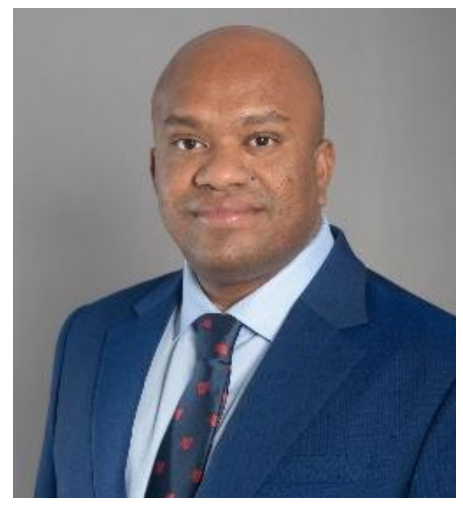

**SURESH RAJA NEETHIRAJAN** is currently a Professor and the University Research Chair of Dalhousie University, Canada, with cross appointments at the Faculty of Computer Science and the Faculty of Agriculture. His interdisciplinary research integrates artificial intelligence, precision agriculture, and digital livestock systems to advance animal welfare and sustainable farming practices. He has previously held academic positions at Wageningen University and Research in The Netherlands and the University of Guelph, Canada. His work focuses on translating advanced AI technologies into practical, scalable solutions at the intersection of agriculture, computing, and environmental sustainability.